\documentclass{IEEEoj}

\usepackage{caption} 
\usepackage{soul}
\usepackage{lipsum}
\usepackage{indentfirst}
\usepackage{booktabs}
\let\labelindent\relax
\usepackage{enumitem}
\usepackage{tabularx}
\usepackage{multirow}
\usepackage{graphicx}
\usepackage{colortbl}
\usepackage{makecell} 
\usepackage[table,xcdraw,dvipsnames]{xcolor}
\usepackage{environ}         
\usepackage{etoolbox}        
\usepackage{pifont}
\usepackage{makecell}
\usepackage{array}
\usepackage{hhline}

\usepackage{cite}

\usepackage{graphicx}
\usepackage{wrapfig}
\usepackage[labelformat=simple]{subcaption}

\usepackage{float}
\usepackage{eurosym}
\usepackage{gensymb}
\usepackage{textpos}

\usepackage{amsfonts, eqnarray, amsmath, amssymb}
\usepackage{amsthm}
\usepackage{mathtools}
\usepackage{xfrac}
\usepackage{breqn}
\usepackage{bm}

\usepackage{algorithm}
\usepackage{algpseudocode}
\usepackage{etoolbox}
\algnewcommand{\LeftComment}[1]{\Statex \(\triangleright\) #1}

\makeatletter
\newcommand*{\algrule}[1][\algorithmicindent]{%
  \hspace*{.2em}
  \vrule 
  \hspace*{\dimexpr#1-.2em-.4pt}%
}

\newcommand{\StatePar}[1]{%
  \State\parbox[t]{\dimexpr\linewidth-\ALG@thistlm}{\strut #1\strut}%
}
\renewcommand{\ALG@beginalgorithmic}{\offinterlineskip}

\newcount\ALG@printindent@tempcnta
\def\ALG@printindent{%
  \ifnum \theALG@nested > 0
    \ifx\ALG@text\ALG@x@notext
    \else
      \unskip
      \ALG@printindent@tempcnta=1
      \loop
        \algrule[\csname ALG@ind@\the\ALG@printindent@tempcnta\endcsname]%
        \advance \ALG@printindent@tempcnta 1
        \ifnum \ALG@printindent@tempcnta<\numexpr\theALG@nested+1\relax
      \repeat
        \fi
    \fi
}
\patchcmd{\ALG@doentity}{\noindent\hskip\ALG@tlm}{\ALG@printindent}{}{\errmessage{failed to patch}}
\makeatother

\algrenewcommand\algorithmicend{\strut\textbf{end}}
\algrenewcommand\algorithmicdo{\strut\textbf{do}}
\algrenewcommand\algorithmicwhile{\strut\textbf{while}}
\algrenewcommand\algorithmicfor{\strut\textbf{for}}
\algrenewcommand\algorithmicforall{\strut\textbf{for all}}
\algrenewcommand\algorithmicloop{\strut\textbf{loop}}
\algrenewcommand\algorithmicrepeat{\strut\textbf{repeat}}
\algrenewcommand\algorithmicuntil{\strut\textbf{until}}
\algrenewcommand\algorithmicprocedure{\strut\textbf{procedure}}
\algrenewcommand\algorithmicfunction{\strut\textbf{function}}
\algrenewcommand\algorithmicif{\strut\textbf{if}}
\algrenewcommand\algorithmicthen{\strut\textbf{then}}
\algrenewcommand\algorithmicelse{\strut\textbf{else}}

\algrenewcommand\algorithmicrequire{\strut\textbf{Input:}}
\algrenewcommand\algorithmicensure{\strut\textbf{Output:}}

\let\oldState\State
\renewcommand{\State}{\oldState\strut}

\usepackage{array}

\usepackage{url}

\NewEnviron{resizealign}{\sbox0{
    $\begin{matrix}\displaystyle\BODY\end{matrix}$}%
  \sbox1{$(\theequation)$}%
  \sbox2{\parbox{\dimexpr \wd0 + 2\wd1}%
    {\begin{align}\BODY\end{align}}}
  \noindent\resizebox{\columnwidth}{!}{\usebox2}%
}

\def\BibTeX{{\rm B\kern-.05em{\sc i\kern-.025em b}\kern-.08em
    T\kern-.1667em\lower.7ex\hbox{E}\kern-.125emX}}
\AtBeginDocument{\definecolor{ojcolor}{cmyk}{0.93,0.59,0.15,0.02}}
\def\OJlogo{\vspace{-4pt}\hskip-4pt\includegraphics[height=18pt]{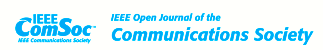}}

\begin{document}
\receiveddate{XX Month, XXXX}
\reviseddate{XX Month, XXXX}
\accepteddate{XX Month, XXXX}
\publisheddate{XX Month, XXXX}
\currentdate{}
\doiinfo{OJCOMS.2026.011100}

\title{Coarse-Grained Predictive Resource Allocation Framework enabling In-Premise XR Collaborations}

\author{Sourav Mondal\IEEEauthorrefmark{1} \IEEEmembership{(Member, IEEE)}, AND Elaine Wong\IEEEauthorrefmark{1}
\IEEEmembership{(Senior Member, IEEE)}}
\affil{Department of Electrical and Electronic Engineering, The University of Melbourne, Parkville, VIC 3010, Australia}
\corresp{CORRESPONDING AUTHOR: Sourav Mondal (e-mail: sourav.mondal@unimelb.edu.au).}
\markboth{Preparation of Papers for IEEE OPEN JOURNALS}{Author \textit{et al.}}

\begin{abstract}
Immersive extended reality (XR) applications are rapidly gaining attention as key enablers of human-centric connectivity over next-generation networks. However, meeting their high-throughput, ultra-low-latency, and high-reliability requirements while preventing cyber-sickness in human users remains a critical challenge, particularly across large geographic spans and in-premise environments with non-deterministic wireless connectivity. Fiber-To-The-Room (FTTR), an extension of the widely deployed Fiber-To-The-Home (FTTH) infrastructure, offers a promising solution; yet satisfying the stringent Quality-of-Experience (QoE) requirements of XR collaborators demands robust resource allocation across the multi-tier cascade of FTTH, FTTR, and WiFi segments. In this paper, we propose a coarse-grained traffic forecasting empowered proactive resource allocation and intelligent in-premise roaming framework where the artificial intelligence (AI) model complexities are selected according to the computational capability of each network segment. At the FTTR segment, we employ a hybrid temporal convolutional and bi-directional long short-term memory (TCN-BiLSTM) network for multi-modal XR traffic forecasting, integrating passive optical network (PON) and WiFi access points with multi-link operation (MLO) and orthogonal frequency-division multiple access (OFDMA). The resulting long-horizon, multi-step-ahead traffic forecasts reduce inference frequency and enable coordinated FTTR-WiFi resource allocation and network-orchestrated seamless user mobility across rooms. At the FTTH segment, we deploy Transformer-based time-series forecasting models, e.g., Informer, Autoformer, Patch Time-Series Transformer (PatchTST), and Temporal Fusion Transformer (TFT) for a more complex multi-tier aggregated traffic. Our numerical evaluations show that Transformer-based models are up to $\sim$75\% more accurate in long-horizon traffic forecasting than the hybrid model. Furthermore, our extensive simulations establish that the proposed framework significantly outperforms traditional resource allocation methods in satisfying the QoE requirements of XR collaborations.
\end{abstract}


\maketitle
\section{INTRODUCTION} \label{sec1}
\begin{figure}[!t]
\centering
\includegraphics[width=\columnwidth,keepaspectratio]{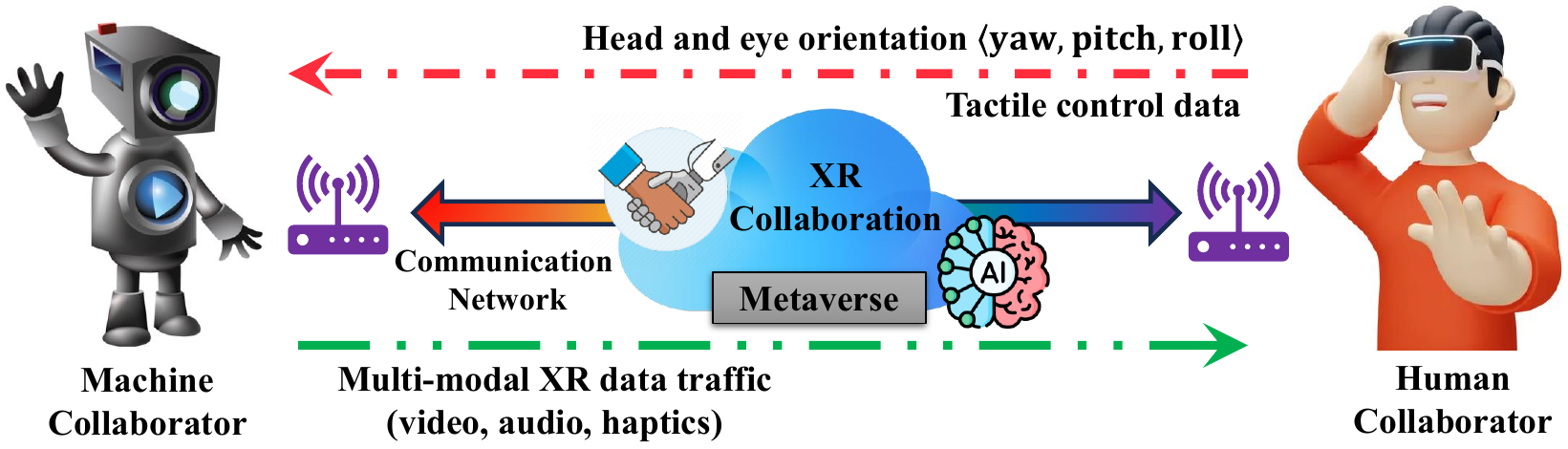}
\caption{A schematic illustration of XR collaboration over networks.}
\label{xrc}
\end{figure}

\IEEEPARstart{O}{riginating} from the entertainment and gaming industries, immersive extended reality (XR) applications such as virtual reality (VR), augmented reality (AR), and mixed reality (MR) are now the primary impetus for future-generation mobile and fiber-wireless networks to expand into new vertical sectors such as automotive, energy, agriculture, healthcare, and manufacturing~\cite{xr_survey}. Further support from cyber-physical systems, edge-cloud servers, and artificial intelligence (AI) potentially empowers XR technologies to contribute to the emerging \emph{Metaverse} paradigm~\cite{ind_metav} which enables human-to-human interactions in remote skill transfer, edutainment, and exergaming, as well as human-to-machine collaborations in smart manufacturing, e-healthcare, and teleoperations~\cite{h2m_sourav}. However, it is essential to note that human users engaged in prolonged XR collaborations may experience \emph{cyber-sickness} if the perceptual continuity of visual and sensory information, with tight synchronization, is not maintained~\cite{xr_iotj}. In other words, multi-modal video, head orientations, tactile control, and haptic traffic must be transmitted within their respective latency and jitter bounds~\cite{xr_std}. For example, achieving an \emph{ideal QoE} requires 16K video (200-600 Mbps) at a frame rate 60~fps to be transmitted within 10~msec latency demanding $\geq$1~Gbps throughput and $\geq$99.999\% reliability~\cite{itu_fttr}. In contrast, head movement, tactile control, and haptic traffic involve only a few Kbps datarate but require transmission within a stricter 1-10 msec latency and jitter. This implies that \emph{ensuring a highly immersive QoE during XR collaborations}, as shown in Fig.~\ref{xrc}, poses a significant resource allocation challenge to network operators as networks must \emph{accommodate aggregated multi-modal XR traffic flows alongside background Internet traffic}. In fact, current market trends indicate that the average number of connected devices per broadband household will rise from 13 in 2020 to 44 by 2030, including millions of XR devices~\cite{wbba2023}.\par
\setlength{\textfloatsep}{5pt}
The challenge of satisfying the multi-dimensional requirements of high-throughput, low-latency, and high-reliability is significantly exacerbated when XR collaborations are deployed across: (a) large geographical spans; and/or (b) in-premise scenarios such as homes and factories, where connectivity often relies on low-datarate, best-effort, and high-latency technologies such as WiFi with carrier-sense multiple access \cite{fttr_shen}. Although passive optical network (PON), commonly deployed in Fiber-To-The-x (FTTx) architecture with a typical reach of 10-20 km, addresses the first challenge through AI-based predictive and adaptive dynamic bandwidth allocation (DBA)~\cite{ml_ewon}, the second challenge remains largely unaddressed. To overcome this bottleneck, the research community is advancing toward the Sixth Generation Fixed Network (F6G) vision, which envisions ubiquitous AI-native intelligence, extreme deterministic latency, and seamless physical-digital fusion across all in-premises network segments~\cite{F6G}. Furthermore, a novel fiber-based reference architecture, known as Fiber-To-The-Room (FTTR), has been standardized as an extension of Fiber-To-The-Home (FTTH) and represents a practical architectural realization aligned with the F6G vision~\cite{fttr_std}.\par
\begin{figure*}[!t]
\centering
\includegraphics[width=\textwidth,keepaspectratio]{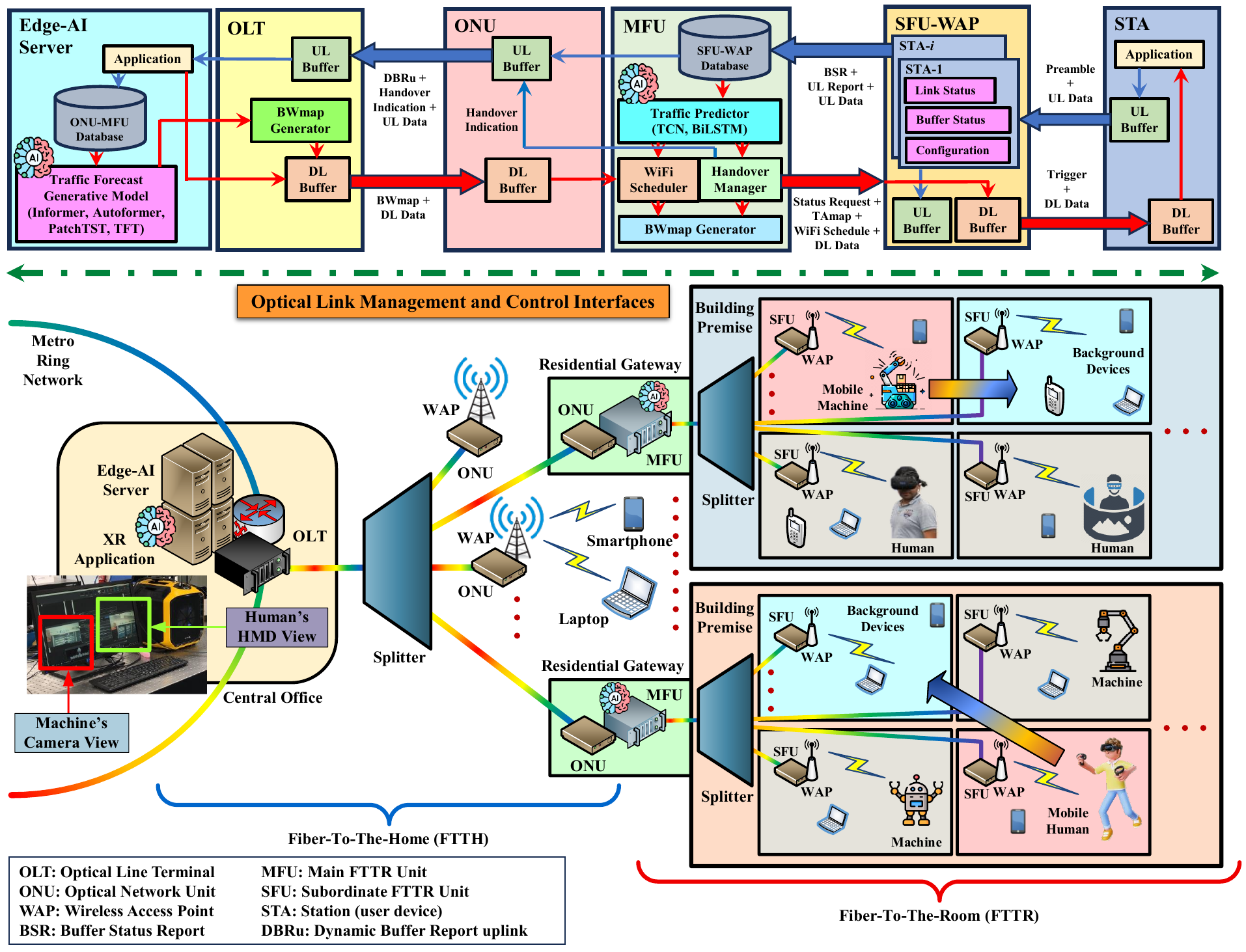}
\caption{The multi-tier integrated FTTR network architecture embedded with coarse-grained traffic forecast AI models. This architecture enables proactive resource allocation and intelligent roaming to satisfy the QoE requirements of multi-modal traffic during XR collaborations.}
\label{arch}
\end{figure*}

Catering to the surging bandwidth demands from future applications like XR collaborations, the International Telecommunication Union’s Telecommunication Standardization Sector (ITU-T) has standardized the 50G-PON supporting a 50 Gbps datarate on a single wavelength (ITU-T G.9804.3) \cite{rene}. Additionally, ITU-T Q3/SG15 initiated the standardization of fiber-based in-premise networks under G.fin series that supports deployment of \emph{internal FTTR segments} cascaded with \emph{external FTTH segments}, as shown in Fig.~\ref{arch}. More specifically, the optical line terminal (OLT) of the external FTTH segment connects multiple optical network units (ONUs) via a 50G-PON. Each ONU may further connect to an internal FTTR segment where a main FTTR unit (MFU) is connected to multiple subordinate FTTR units (SFUs) via an XG(S)-PON supporting a 10 Gbps datarate. Interestingly, SFUs can be further integrated with mobile 5G/6G, WiFi-6/6E/7/8, mmWave, THz, or LiFi wireless access points (WAPs) in each room to provide wireless connectivity to user devices or stations (STAs)~\cite{fttr_mag}. It is noteworthy that MFUs act as \emph{residential gateways}, relaying data between FTTH and FTTR segments while also serving as centralized controllers for the XG(S)-PON and the associated WAPs. Moreover, by leveraging new control and management interfaces, MFUs enable coordinated FTTR-WiFi resource allocation, intelligent user roaming across rooms, interference mitigation and power optimization~\cite{fttr_zte}.\par
Complementing fiber technologies, WiFi standards are also being engineered to support the stringent QoE requirements of applications like XR collaborations. Theoretically, WiFi-6/6E supports a maximum datarate of $\sim$9.6 Gbps with best-case latency of $\sim$2 msec, WiFi-7 increases this to $\sim$46 Gbps with 1-2 msec latency, and WiFi-8 is being standardized to support 100+~Gbps with sub-millisecond latency~\cite{wifi7}. In addition to supporting high-throughput with low-latency, each standard also introduces significant functional and architectural enhancements. For example, WiFi-6 adopts uplink scheduling and orthogonal frequency-division multiple access (OFDMA) to enhance spectrum efficiency and reduce latency. Building on this, WiFi-7 introduces multi-link operation (MLO), enabling devices to aggregate multiple frequency bands for a higher throughput. The forthcoming WiFi-8 is focused on coordinated multi-WAP transmission, aiming to minimize interference and ensure deterministic, ultra-low-latency communication. Nevertheless, \emph{satisfaction of QoE requirements of multi-modal traffic in XR collaborations aggregated with background traffic over FTTR-WiFi architecture, especially in dense multi-WAP in-premise scenarios is yet to be investigated.}\par
\setlength{\textfloatsep}{1pt}
Note that distributed WiFi networks lack received-power-threshold-based handover mechanisms. Even with poorly received power, a STA can decode beacons with low-order modulation and continues to be connected to a WAP despite being unable to transmit or receive any data. Once the STA is disconnected from a WAP, another connection with a new WAP and a network path to the application server is established, thus \emph{interrupting seamless data transmission}. In \cite{fttrb_ofc}, we proposed an intelligent in-premise roaming framework for XR collaborations exploiting the centralized FTTR-WiFi architecture. However, a detailed formulation of the network-orchestrated global handovers and thorough evaluation of the proposed solution is still pending.\par
In practical deployments, XR collaboration platforms are typically hosted on edge-cloud servers co-located with the optical line terminal (OLT), commonly referred to as \emph{Edge-AI servers}. Hence, traffic from multiple in-premise STAs are first aggregated at the WAPs/SFUs over the wireless interface, and then traffic from multiple SFUs are further aggregated at MFUs/ONUs over the internal FTTR, to be transmitted to the OLT over the external FTTH \cite{fttr-b}. We critically observe that employing traditional PON and WiFi resource allocation schemes across the multi-tier cascade of FTTH, FTTR, and WiFi or the integrated FTTR architecture is highly inefficient to guarantee the QoE requirements of XR collaboration as they are non-coordinated and reactive to the traffic load in nature \cite{g_shen}. These pressing issues motivate us to propose AI-empowered traffic forecasting for proactive resource allocation and intelligent in-premise roaming.\par
Within internal FTTR segments, MFUs forecast future traffic from STAs associated with each WAP based on collected network information via FTTR control and management interfaces. Based on these forecasts, WAPs are notified of their wireless bandwidth allocation to associated STAs in upcoming timeslots and the corresponding time-assignment map (TAmap) for SFUs are also generated \cite{ITUTG9942}. In addition, a handover process is also initiated for STAs whose QoE are degrading. The inherent long-term dependencies and complex pseudo-periodic, bursty, and heavy-tailed patterns of multi-modal XR traffic \cite{xr_iotj} necessitate the use of sophisticated yet lightweight forecasting models, given the limited computational resources at MFUs, such as Temporal Convolutional Networks (TCN) and Bi-directional Long Short-Term Memory (BiLSTM) networks. Thus, combining the translation-invariant local pattern extraction ability of TCN with the bi-directional long-range dependency modeling capability of BiLSTM, we design a hybrid TCN-BiLSTM network with 1~TCN and 2~BiLSTM layers to achieve highly accurate traffic forecasts.\par
At the ONUs of external FTTH segment, the arriving multi-tier multi-modal aggregated traffic exhibits highly complex and dynamic characteristics. Consequently, the accuracy of the aforementioned predictive models deteriorates, especially for long-horizon forecasting. As the polling cycles of 50G-PON in the FTTH segment follow the standard 125 $\mu$sec polling cycle duration \cite{gpon}, making long-horizon traffic forecasts at a coarser interval of 1-10 msec for multi-samples-ahead bandwidth map (BWmap) generation can significantly reduce \emph{computational resource demands}. Hence, we adopt \emph{Transformer-based time-series forecast models}, namely Informer, Autoformer, Patch Time-Series Transformer (PatchTST), and Temporal Fusion Transformer (TFT) \cite{transformers_mdpi}. With approximately 1-5 million parameters, these models are significantly lighter than popular large-language models with billion/trillion-parameters. While their offline training phase demands substantial computational resources, their highly parallelizable architecture can be leveraged for generating rapid inference necessary for real-time network deployment. They interpret both short-term dynamics and long-term temporal dependencies arising due to pseudo-periodic bursts in the historical traffic data under different network load conditions by harnessing \emph{different variations of self-attention mechanisms}~\cite{genai}. A preliminary version of this work showcasing our 25G-PON testbed emulating the FTTH segment was presented in \cite{fttr_ofc_26}. This paper significantly extends the prior work by evaluating the performance of Transformer-based models in long-horizon forecasting of aggregated multi-modal traffic. We also establish through extensive network simulations that our proposed framework satisfies the QoE requirements of XR collaborations. Our primary contributions are:
\begin{enumerate}[label=(\roman*)]
    \item We design an multi-tier integrated FTTR network architecture embedded with neural network and transformer-based AI models for proactive resource allocation with robustness against prediction errors.
    \item We propose a jointly coordinated FTTR-WiFi predictive resource allocation scheme where WiFi-7 WAPs are connected by a XG(S)-PON. To dynamically assign bandwidth to STAs by WAPs with OFDMA and MLO in the 5/6~GHz bands, we formulate an optimization problem and devise a time-efficient solution algorithm.
    \item We verify that our centralized WiFi-STA resource allocation scheme outperforms standard dense WiFi distributed systems to keep the air-interface latency of XR multi-modal traffic $\leq$10 msec even under high-load conditions.
    \item We observe that the proposed intelligent in-premise seamless roaming reduces air-interface latency up to $\sim$90\% through network-orchestrated handovers as XR collaborators move across multiple rooms.
    \item We assess the normalized root mean squared error (NRMSE) of the considered predictive AI-based models to predict the multi-modal traffic during XR collaborations over FTTR-WiFi segment. We observe that the best NRMSE is achieved by the hybrid 1~TCN+2~BiLSTM model under all network load conditions.
    \item We compare the training and inference times of the Transformer-based models and evaluate their forecast accuracies for the multi-tier aggregated traffic in the FTTH segment. Results show that these models are $\sim$75\% more accurate than the neural network-based models against a forecast horizon of up to 80 samples.
\end{enumerate}
\par The remainder of this paper is organized as follows. Section~\ref{sec2} reviews recent related literature. Section~\ref{sec3} describes XR collaborations and characterizes the associated multi-modal traffic. Section~\ref{sec4} details the interfaces of the multi-tier integrated FTTR network architecture, with emphasis on the newly introduced control and management functions in the FTTR segment. Section~\ref{sec5} introduces the proposed coarse-grained traffic-predictive framework for resource allocation as well as develops the intelligent in-premise roaming mechanism. Section~\ref{sec6} evaluates the performance of the proposed framework in practical usage scenarios. Finally, Section~\ref{sec7} summarizes the key insights from this research.

\begin{table*}[!t]
\caption{Summary of QoE requirements for immersive XR applications \cite{xr_std,xr_iotj}}
\label{table1}
\centering
%
\setlength\arrayrulewidth{1.5pt}
\begin{tabular}{
>{\columncolor[HTML]{EFEFEF}}c 
>{\columncolor[HTML]{EFEFEF}}c cccccc}
\hline
\multicolumn{8}{c}{\cellcolor[HTML]{C0C0C0}\textbf{Traffic Profile}} \\ \hline
\multicolumn{2}{c|}{\cellcolor[HTML]{EFEFEF}} &
  \multicolumn{1}{c|}{\cellcolor[HTML]{EFEFEF}} &
  \multicolumn{1}{c|}{\cellcolor[HTML]{EFEFEF}} &
  \multicolumn{1}{c|}{\cellcolor[HTML]{EFEFEF}} &
  \multicolumn{1}{c|}{\cellcolor[HTML]{EFEFEF}} &
  \multicolumn{1}{c|}{\cellcolor[HTML]{EFEFEF}} &
  \cellcolor[HTML]{EFEFEF} \\
\multicolumn{2}{c|}{\multirow{-2}{*}{\cellcolor[HTML]{EFEFEF}\textbf{Traffic Type}}} &
  \multicolumn{1}{c|}{\multirow{-2}{*}{\cellcolor[HTML]{EFEFEF}\textbf{\begin{tabular}[c]{@{}c@{}}Frame\\ Rate\end{tabular}}}} &
  \multicolumn{1}{c|}{\multirow{-2}{*}{\cellcolor[HTML]{EFEFEF}\textbf{\begin{tabular}[c]{@{}c@{}}Avg.\\ Datarate\end{tabular}}}} &
  \multicolumn{1}{c|}{\multirow{-2}{*}{\cellcolor[HTML]{EFEFEF}\textbf{Bandwidth}}} &
  \multicolumn{1}{c|}{\multirow{-2}{*}{\cellcolor[HTML]{EFEFEF}\textbf{\begin{tabular}[c]{@{}c@{}}Packet\\ Arrival Time\end{tabular}}}} &
  \multicolumn{1}{c|}{\multirow{-2}{*}{\cellcolor[HTML]{EFEFEF}\textbf{\begin{tabular}[c]{@{}c@{}}Packet\\ Size (Byte)\end{tabular}}}} &
  \multirow{-2}{*}{\cellcolor[HTML]{EFEFEF}\textbf{\begin{tabular}[c]{@{}c@{}}Latency\\ Bound\end{tabular}}} \\ \hline
\multicolumn{1}{c|}{\cellcolor[HTML]{EFEFEF}} &
  \multicolumn{1}{c|}{\cellcolor[HTML]{EFEFEF}} &
  \multicolumn{1}{c|}{} &
  \multicolumn{1}{c|}{} &
  \multicolumn{1}{c|}{} &
  \multicolumn{1}{c|}{} &
  \multicolumn{1}{c|}{} &
   \\
\multicolumn{1}{c|}{\cellcolor[HTML]{EFEFEF}} &
  \multicolumn{1}{c|}{\multirow{-2}{*}{\cellcolor[HTML]{EFEFEF}\textbf{\begin{tabular}[c]{@{}c@{}}Fair QoE\\ {[}4K{]}\end{tabular}}}} &
  \multicolumn{1}{c|}{\multirow{-2}{*}{\begin{tabular}[c]{@{}c@{}}60\\ fps\end{tabular}}} &
  \multicolumn{1}{c|}{\multirow{-2}{*}{\begin{tabular}[c]{@{}c@{}}40-90\\ Mbps\end{tabular}}} &
  \multicolumn{1}{c|}{\multirow{-2}{*}{\begin{tabular}[c]{@{}c@{}}$\geq$ 260\\ Mbps\end{tabular}}} &
  \multicolumn{1}{c|}{} &
  \multicolumn{1}{c|}{} &
  \multirow{-2}{*}{$\leq$20 msec} \\ \cline{2-5} \cline{8-8} 
\multicolumn{1}{c|}{\cellcolor[HTML]{EFEFEF}} &
  \multicolumn{1}{c|}{\cellcolor[HTML]{EFEFEF}\textbf{\begin{tabular}[c]{@{}c@{}}Comfortable\\ QoE {[}8K{]}\end{tabular}}} &
  \multicolumn{1}{c|}{\begin{tabular}[c]{@{}c@{}}{[}60,90{]}\\ fps\end{tabular}} &
  \multicolumn{1}{c|}{\begin{tabular}[c]{@{}c@{}}100-360\\ Mbps\end{tabular}} &
  \multicolumn{1}{c|}{\begin{tabular}[c]{@{}c@{}}$\geq$ 1\\ Gbps\end{tabular}} &
  \multicolumn{1}{c|}{} &
  \multicolumn{1}{c|}{} &
  $\leq$15 msec \\ \cline{2-5} \cline{8-8} 
\multicolumn{1}{c|}{\multirow{-4}{*}{\cellcolor[HTML]{EFEFEF}\textbf{Video}}} &
  \multicolumn{1}{c|}{\cellcolor[HTML]{EFEFEF}\textbf{\begin{tabular}[c]{@{}c@{}}Ideal QoE\\ {[}16K{]}\end{tabular}}} &
  \multicolumn{1}{c|}{\begin{tabular}[c]{@{}c@{}}{[}60,90,120{]}\\ fps\end{tabular}} &
  \multicolumn{1}{c|}{\begin{tabular}[c]{@{}c@{}}400-800\\ Mbps\end{tabular}} &
  \multicolumn{1}{c|}{\begin{tabular}[c]{@{}c@{}}$\geq$ 1.5\\ Gbps\end{tabular}} &
  \multicolumn{1}{c|}{\multirow{-4}{*}{\begin{tabular}[c]{@{}c@{}}Truncated Gaussian,\\ Mean = $\frac{1}{\text{fps}}$,\\ SD = 2 msec\end{tabular}}} &
  \multicolumn{1}{c|}{\multirow{-4}{*}{\begin{tabular}[c]{@{}c@{}}Truncated Gaussian,\\ Mean = $\frac{\text{Avg. Datarate}}{(\text{fps} \times 8)}$,\\ SD = 10.5\% of Mean\end{tabular}}} &
  $\leq$10 msec \\ \hline
\multicolumn{2}{c|}{\cellcolor[HTML]{EFEFEF}\textbf{\begin{tabular}[c]{@{}c@{}}HMD \\ Orientations\end{tabular}}} &
  \multicolumn{1}{c|}{66 fps} &
  \multicolumn{1}{c|}{\begin{tabular}[c]{@{}c@{}}0.2-0.3\\ Mbps\end{tabular}} &
  \multicolumn{1}{c|}{\begin{tabular}[c]{@{}c@{}}$\geq$ 50\\ Mbps\end{tabular}} &
  \multicolumn{1}{c|}{\begin{tabular}[c]{@{}c@{}}Gamma,\\ Mean = 15 msec,\\ SD = 2 msec\end{tabular}} &
  \multicolumn{1}{c|}{100-500} &
  $\leq$10 msec \\ \hline
\multicolumn{2}{c|}{\cellcolor[HTML]{EFEFEF}\textbf{\begin{tabular}[c]{@{}c@{}}Tactile \\ Control\end{tabular}}} &
  \multicolumn{1}{c|}{100 fps} &
  \multicolumn{1}{c|}{\begin{tabular}[c]{@{}c@{}}1-2 \\ Mbps\end{tabular}} &
  \multicolumn{1}{c|}{\begin{tabular}[c]{@{}c@{}}$\geq$ 200\\ Mbps\end{tabular}} &
  \multicolumn{1}{c|}{\begin{tabular}[c]{@{}c@{}}Generalized Pareto,\\ Mean = 10 msec,\\ SD = 2 msec\end{tabular}} &
  \multicolumn{1}{c|}{1000-1500} &
  1-10 msec \\ \hline
\multicolumn{2}{c|}{\cellcolor[HTML]{EFEFEF}\textbf{\begin{tabular}[c]{@{}c@{}}Haptic \\ Feedback\end{tabular}}} &
  \multicolumn{1}{c|}{100 fps} &
  \multicolumn{1}{c|}{\begin{tabular}[c]{@{}c@{}}1-2\\ Mbps\end{tabular}} &
  \multicolumn{1}{c|}{\begin{tabular}[c]{@{}c@{}}$\geq$ 200\\ Mbps\end{tabular}} &
  \multicolumn{1}{c|}{\begin{tabular}[c]{@{}c@{}}Generalized Pareto,\\ Mean = 10 msec,\\ SD = 2 msec\end{tabular}} &
  \multicolumn{1}{c|}{1000-1500} &
  1-10 msec \\ \hline
\end{tabular}
\end{table*}
\setlength{\textfloatsep}{1pt}

\section{Review of Related Works} \label{sec2}
Deploying XR collaborations over existing communication networks, especially across large geographic spans and in-premise environments with non-deterministic wireless connectivity, presents several complex challenges to the researchers. Table~\ref{table1} summarizes the \emph{fair}, \emph{comfortable}, and \emph{ideal QoE} requirements associated with multi-modal XR traffic, encompassing video, head movement, tactile control, and haptic data. It is also evident from this table that ensuring ultra-low-latency with low-jitter are highly essential alongside high-throughput. To satisfy these requirements, application layer and network layer collaborative approaches like network-aware adaptive XR video streaming \cite{xr_5g1} and XR QoE-aware scheduling \cite{xr_5g2} are proposed. Moreover, the authors of \cite{vr_mpg} proposed a framework for immersive multi-user game where XR video frames of each player according to their field-of-view (FOV) and game status are pre-rendered for low-latency transmission. The authors of \cite{lstm_xr} proposed a long short-term memory (LSTM)-based FOV prediction method to transmit the FOV tiles of XR frames with highest quality and the rest with lower quality to reduce 61\% bandwidth demand. Nonetheless, these solutions alone are incompatible to satisfy the QoE requirements of XR collaborations without increasing network capacity and ensuring deterministic connectivity.\par
In \cite{xr_apc}, we explored the potential of integrated FTTR architecture to meet these criteria. As FTTH networks are widely deployed with the standard ONU management and control interface (OMCI) and FTTR networks are conceptualized as an extension of FTTH, the enhanced OMCI (eOMCI) is further standardized for transparently forwarding control messages over OLT-MFU and MFU-SFU interfaces \cite{fttr_ofc_25}. Note that in traditional FTTR architecture, the medium access control (MAC) layers of PON and WiFi operate independently. To improve the inter-MAC coordination, the authors of \cite{fttr_jocn} proposed a centralized FTTR-WiFi integrated network (FWIN) that satisfy the \emph{telecom-quality WiFi experience} and provide with premium home broadband connectivity. Using field-trial measurements, they demonstrated the superiority of centralized FWIN over segmented networks for resource allocation, interference avoidance, and latency reduction. Similarly, the authors of \cite{ponizing_wifi} proposed another unified PON and WiFi MAC protocol, termed ``PONizing WiFi," to enhance network access rates, resource utilization, and handover times. Additionally, the authors of \cite{green_fttr} analyzed the power consumption of FTTR deployments to propose an efficient energy-saving mechanism. These seminal works highlight the architectural and energy-efficiency benefits of integrated FTTR-WiFi architecture.\par
Recently introduced OFDMA in WiFi-6/6E and MLO in WiFi-7 significantly mitigate intra-WAP channel contention and enhance spectral efficiency. The authors of \cite{wifi6_sim} evidenced throughput and latency variations of WiFi-6 on a real experimental testbed against signal-to-noise-ratio (SNR), packet size, buffer window size, channel occupancy, to name a few. Furthermore, the authors of \cite{wifi6_sch} designed efficient algorithms for optimal assignment of subcarriers to user groups while maximizing aggregated throughput and the authors of \cite{wifi7_sch} proposed an intelligent channel allocation framework for WiFi-7 MLO networks using online learning to enhance spectrum utilization. Although these solutions greatly reduce air-interface latency for static STAs, neither of them inherently support seamless inter-WAP handovers in dense WiFi deployments. Consequently, the QoE requirements of mobile STAs involved in immersive XR collaborations are not satisfied when the channel conditions become poor. Nevertheless, the centralized FTTR-WiFi integrated architecture can be leveraged to enable intelligent network-orchestrated in-premise roaming \cite{fttr_coop}. The authors of \cite{fttr_jocn} experimentally demonstrated seamless inter-WAP handovers between 4$\times$4~meter$^2$ rooms with $\leq$20 msec latency. In addition, the authors of \cite{fttr_ofc_25} performed the a field-trial of inter-WAP handovers with $\leq$35 msec latency where the WAPs were 6~meter apart. We also performed network simulations in \cite{fttrb_ofc} to show that efficient network-orchestrated handovers between WAPs in 10$\times$10~meter$^2$ rooms can reduce air-interface latency up to 90\%.\par

The ability of the integrated FTTR-WiFi architecture to meet stringent QoE requirements of XR collaborations can be further enhanced through proactive resource allocation. In \cite{xr_iotj}, we proposed an efficient end-to-end resource allocation scheme based on user head-movement predictions to enable XR collaborations between remote industrial premises. However, such user-sensitive application data is often unavailable for collection from real networks. Hence, employing traffic prediction models based on historical traffic traces is essential in designing proactive resource allocation frameworks. Interestingly, the authors of \cite{cnn_lstm} showed impressive traffic prediction accuracy on open-source Zenodo and PeMS datasets using convolutional neural networks (CNNs), recurrent neural networks (RNNs), LSTM networks, and their hybrid combinations. The authors of \cite{trans_gen} designed a generative AI-based traffic forecasting model by integrating a graph neural network with a Transformer. Focusing on traffic in optical metro and access networks, the authors of \cite{jlt_gen_ai} proposed a novel spatio-temporal Transformer that integrates graph attention mechanisms to capture spatial correlations and a Transformer encoder to effectively model long-range temporal dependencies. In addition, the authors of \cite{fttr_genai} introduced a traffic prediction model integrating a Crested Porcupine optimizer, bi-directional-TCN, and BiLSTM networks to accurately  predict complex FTTR traffic. Although these foundational studies show great promise in making highly-accurate traffic predictions, their performance on multi-tier aggregated multi-modal XR traffic over the integrated FTTR architecture warrants thorough investigation.

\begin{table*}[!t]
\caption{Summary of multi-modal traffic in immersive XR applications}
\label{table2}
\centering
%
\setlength\arrayrulewidth{1.5pt}
\begin{tabular}{
>{\columncolor[HTML]{EFEFEF}}c cccccc}
\hline
\multicolumn{7}{c}{\cellcolor[HTML]{CBCEFB}\textbf{XR Application}} \\ \hline
\multicolumn{1}{c|}{\cellcolor[HTML]{EFEFEF}} &
  \multicolumn{1}{c|}{\cellcolor[HTML]{ECF4FF}} &
  \multicolumn{1}{c|}{\cellcolor[HTML]{ECF4FF}} &
  \multicolumn{1}{c|}{\cellcolor[HTML]{ECF4FF}} &
  \multicolumn{1}{c|}{\cellcolor[HTML]{ECF4FF}} &
  \multicolumn{2}{c}{\cellcolor[HTML]{ECF4FF}\textbf{H2M}} \\ \cline{6-7} 
\multicolumn{1}{c|}{\multirow{-2}{*}{\cellcolor[HTML]{EFEFEF}\textbf{Traffic Type}}} &
  \multicolumn{1}{c|}{\multirow{-2}{*}{\cellcolor[HTML]{ECF4FF}\textbf{VR}}} &
  \multicolumn{1}{c|}{\multirow{-2}{*}{\cellcolor[HTML]{ECF4FF}\textbf{AR}}} &
  \multicolumn{1}{c|}{\multirow{-2}{*}{\cellcolor[HTML]{ECF4FF}\textbf{MR}}} &
  \multicolumn{1}{c|}{\multirow{-2}{*}{\cellcolor[HTML]{ECF4FF}\textbf{H2H}}} &
  \multicolumn{1}{c|}{\cellcolor[HTML]{ECF4FF}\textbf{Human}} &
  \cellcolor[HTML]{ECF4FF}\textbf{Machine} \\ \hline
\multicolumn{1}{c|}{\cellcolor[HTML]{EFEFEF}} &
  \multicolumn{1}{c|}{} &
  \multicolumn{1}{c|}{} &
  \multicolumn{1}{c|}{} &
  \multicolumn{1}{c|}{} &
  \multicolumn{1}{c|}{} &
   \\
\multicolumn{1}{c|}{\cellcolor[HTML]{EFEFEF}} &
  \multicolumn{1}{c|}{} &
  \multicolumn{1}{c|}{} &
  \multicolumn{1}{c|}{} &
  \multicolumn{1}{c|}{} &
  \multicolumn{1}{c|}{} &
   \\
\multicolumn{1}{c|}{\cellcolor[HTML]{EFEFEF}} &
  \multicolumn{1}{c|}{} &
  \multicolumn{1}{c|}{} &
  \multicolumn{1}{c|}{} &
  \multicolumn{1}{c|}{} &
  \multicolumn{1}{c|}{} &
   \\
\multicolumn{1}{c|}{\multirow{-4}{*}{\cellcolor[HTML]{EFEFEF}\textbf{Video}}} &
  \multicolumn{1}{c|}{\multirow{-4}{*}{DL}} &
  \multicolumn{1}{c|}{\multirow{-4}{*}{\begin{tabular}[c]{@{}c@{}}DL/\\ UL\end{tabular}}} &
  \multicolumn{1}{c|}{\multirow{-4}{*}{\begin{tabular}[c]{@{}c@{}}DL/\\ UL\end{tabular}}} &
  \multicolumn{1}{c|}{\multirow{-4}{*}{\begin{tabular}[c]{@{}c@{}}DL/\\ UL\end{tabular}}} &
  \multicolumn{1}{c|}{\multirow{-4}{*}{DL}} &
  \multirow{-4}{*}{UL} \\ \hline
\multicolumn{1}{c|}{\cellcolor[HTML]{EFEFEF}\textbf{\begin{tabular}[c]{@{}c@{}}HMD \\ Orientations\end{tabular}}} &
  \multicolumn{1}{c|}{UL} &
  \multicolumn{1}{c|}{UL} &
  \multicolumn{1}{c|}{UL} &
  \multicolumn{1}{c|}{UL} &
  \multicolumn{1}{c|}{UL} &
  DL \\ \hline
\multicolumn{1}{c|}{\cellcolor[HTML]{EFEFEF}\textbf{\begin{tabular}[c]{@{}c@{}}Tactile \\ Control\end{tabular}}} &
  \multicolumn{1}{c|}{UL} &
  \multicolumn{1}{c|}{UL} &
  \multicolumn{1}{c|}{UL} &
  \multicolumn{1}{c|}{UL} &
  \multicolumn{1}{c|}{UL} &
  DL \\ \hline
\multicolumn{1}{c|}{\cellcolor[HTML]{EFEFEF}\textbf{\begin{tabular}[c]{@{}c@{}}Haptic \\ Feedback\end{tabular}}} &
  \multicolumn{1}{c|}{DL} &
  \multicolumn{1}{c|}{DL} &
  \multicolumn{1}{c|}{DL} &
  \multicolumn{1}{c|}{DL} &
  \multicolumn{1}{c|}{DL} &
  UL \\ \hline
\multicolumn{1}{c|}{\cellcolor[HTML]{ECF4FF}\textbf{Environment}} &
  \multicolumn{1}{c|}{\begin{tabular}[c]{@{}c@{}}Active\\ Virtual\end{tabular}} &
  \multicolumn{1}{c|}{\begin{tabular}[c]{@{}c@{}}Passive\\ Virtual\\ and Real\end{tabular}} &
  \multicolumn{1}{c|}{\begin{tabular}[c]{@{}c@{}}Active\\ Virtual\\ and Real\end{tabular}} &
  \multicolumn{1}{c|}{\begin{tabular}[c]{@{}c@{}}Active\\ Virtual\end{tabular}} &
  \multicolumn{2}{c}{\begin{tabular}[c]{@{}c@{}}Active\\ Virtual\\ and Real\end{tabular}} \\ \hline
\multicolumn{1}{c|}{\cellcolor[HTML]{ECF4FF}\textbf{Display Device}} &
  \multicolumn{1}{c|}{\begin{tabular}[c]{@{}c@{}}HMD\\ {[}Occlusion{]}\end{tabular}} &
  \multicolumn{1}{c|}{\begin{tabular}[c]{@{}c@{}}OHMD\\ {[}Translucent{]}\end{tabular}} &
  \multicolumn{1}{c|}{\begin{tabular}[c]{@{}c@{}}OHMD\\ {[}Translucent{]}\end{tabular}} &
  \multicolumn{1}{c|}{\begin{tabular}[c]{@{}c@{}}HMD\\ {[}Occlusion{]}\end{tabular}} &
  \multicolumn{2}{c}{\begin{tabular}[c]{@{}c@{}}HMD\\ {[}Occlusion{]}\end{tabular}} \\ \hline
\multicolumn{1}{c|}{\cellcolor[HTML]{ECF4FF}\textbf{Refresh Rate}} &
  \multicolumn{1}{c|}{{[}90,120{]} Hz} &
  \multicolumn{1}{c|}{{[}60,90{]} Hz} &
  \multicolumn{1}{c|}{{[}90,120{]} Hz} &
  \multicolumn{1}{c|}{{[}90,120{]} Hz} &
  \multicolumn{2}{c}{{[}60,90{]} Hz} \\ \hline
\multicolumn{1}{c|}{\cellcolor[HTML]{ECF4FF}\textbf{Usage Scenario}} &
  \multicolumn{1}{c|}{In-premise} &
  \multicolumn{1}{c|}{\begin{tabular}[c]{@{}c@{}}In-premise/\\ Outdoor\end{tabular}} &
  \multicolumn{1}{c|}{\begin{tabular}[c]{@{}c@{}}In-premise/\\ Outdoor\end{tabular}} &
  \multicolumn{1}{c|}{In-premise} &
  \multicolumn{2}{c}{\begin{tabular}[c]{@{}c@{}}In-premise/\\ Outdoor\end{tabular}} \\ \hline
\end{tabular}
\end{table*}
\setlength{\textfloatsep}{1pt}
\section{Immersive XR Collaborations} \label{sec3}
In this section, we discuss the characteristics of single-user immersive XR applications encompassing virtual reality (VR), augmented reality (AR), and mixed reality (MR), as well as their multi-user interactive extensions like human-to-human (H2H) and human-to-machine (H2M) collaborations. As ultra-high-resolution (4K, 8K, 16K) video, head-movement data as head-mounted display (HMD) orientations, tactile control signals, and haptic feedback coexist in each session \cite{xr_std}, understanding the traffic characteristics and corresponding QoE requirements of these modalities is therefore essential for designing proactive network resource allocation schemes, as presented Table~\ref{table2}.\par
%
\subsection{Characteristics of Multi-modal XR Traffic} \label{sec3.1}
%
\begin{itemize}
    \item[\ding{112}] \textbf{Video.} This is the dominant traffic component in each XR session and can be single stream for both eyes combined or multi-stream for left/right eye separated. According to 3GPP XR traffic model \cite{xr_3gpp}, video frames are pseudo-periodically generated at the source, i.e., if $x(t)$ is a \emph{periodic signal} with period $T$, then, $x(t) = x(t + T), \forall t$, but a \emph{pseudo-periodic signal} $y(t)$ with average period $T$ can be defined as $y(t)\approx y(t+T+\epsilon(t)), \forall t$, where $\epsilon(t)$ is a slowly varying function following any arbitrary distribution. If we model the traffic data as a time-series, we will observe that the \emph{trend} is absent, the \emph{seasonality} is represented by the average period $T$, and the \emph{jitter} is represented by $\epsilon(t)$. Typically, the video frame generation period follows a truncated Gaussian distribution with Mean = 1/(frame rate), standard deviation (SD) = 2 msec, and [Min, Max] = $[-4,4]$ msec or $[-5,5]$ msec. In addition, video frame sizes also follow a Truncated Gaussian distribution with Mean = (average datarate)/(frame rate $\times$ 8) bytes and [SD, Min, Max] = $[10.5\%, 50 \%, 150\%]$ of Mean. Note that 4K, 8K, and 16K video frame qualities are mandatory to ensure fair, comfortable, and ideal QoE, respectively. Each frame is further segmented into multiple Ethernet packets with payload $\leq 1500$ bytes and transmitted as a burst over networks. Nevertheless, recent frame compression techniques exploiting temporal redundancy between consecutive frames such as high efficiency video coding (H.265/HEVC) and versatile video coding (H.266/VVC) \cite{vvc} can significantly reduce the datarate when the frames are transmitted with high-reliability. For smoothness and responsiveness of the visuals, modern HMDs/OHMDs typically support a frame refresh rate of  90-120~Hz \cite{deb_vr}.
    \item[\ding{112}] \textbf{HMD orientations.} The orientations and physical positions of an HMD/OHMD play a critical role in XR applications, particularly for enabling optimized viewport-dependent streaming (VDS) instead of resource-intensive $360\degree$ video streaming. As VDS dynamically selects and transmits only the portion of the scene that the user is currently looking at or the ``viewport", the datarate is reduced significantly \cite{vr_mpg}. Usually HMD or optical HMD (OHMD) motion is captured in two formats: three degrees-of-freedom (3DoF), which tracks orientation only, or six degrees-of-freedom (6DoF), which tracks both orientation and physical position, resulting in 100-500 bytes. Orientation can be represented as Euler angles (yaw, pitch, and roll), or as normalized quaternions to avoid issues like gimbal lock. The physical position is tracked as $(x, y, z)$-space coordinates. In \cite{xr_iotj}, based on our experimental observations, we modeled pseudo-periodic HMD orientations traffic inter-arrival time following a Gamma distribution with Mean = 15 msec and SD = 2 msec.
    \item[\ding{112}] \textbf{Tactile control.} While participating in XR applications, humans can manipulate virtual or remote objects using tactile controllers or gloves embedded with orientation and flex sensors. The spatial movements of the hands and fingers are tracked using 3DoF/6DoF representations. Additionally, tension and bending angles across various hand segments measured by the flex sensors are digitized and aggregated into packets of 1000-1500 bytes. According to the experimental data from commercially available devices, the tactile control traffic is modeled as pseudo-periodic with Mean = 10~msec and jitter as a generalized Pareto distribution with SD = 2~msec \cite{h2m_sourav}.
    \item[\ding{112}] \textbf{Haptic feedback.} To replicate human touch and motion, haptic feedback is generated upon impact between the remote machine and physical or virtual objects. This feedback encompasses both \emph{kinesthetic perception}, capturing forces, torques, positions, and velocities perceived through muscles, joints, and tendons, and \emph{tactile perception}, reflecting surface texture, friction, and pressure sensed via mechanoreceptors in the skin. The generated data are aggregated into packets of 1000-1500 bytes. Consistent with experimental data from commercial systems, this haptic traffic is modeled as pseudo-periodic with a mean of 10~msec, and jitter following a generalized Pareto distribution with a standard deviation of 2~msec~\cite{h2m_sourav}.
    
\end{itemize}

\subsection{QoE Requirements of Immersive XR Applications} \label{sec3.2}
\begin{itemize}
    \item[\ding{112}] \textbf{Virtual Reality (VR).} Human users are immersed in a fully computer-generated environment replacing real-world sensory input with synthetic visual, auditory, and haptics. Their HMD orientations and tactile control signals are transmitted in the uplink (UL) to the application server. In turn, their presence in the virtual space and interactions with active virtual objects are enabled. Followed by their viewport-dependent video and haptic feedback are transmitted in the downlink (DL) \cite{vr_bk}. Therefore, the UL traffic from humans only demand low-latency and high-reliability, but the DL traffic requires high-throughput, low-latency, and high-reliability.
    \item[\ding{112}] \textbf{Augmented Reality (AR).} Digital content and passive virtual objects are overlaid onto the physical world in real-time, allowing human users to perceive both experiences simultaneously. Thus, optical see-through or video see-through displays, e.g., OHMDs, smartphones, or smart glasses are required that align virtual objects with the user’s real-world FOV. Note that \emph{split-rendering} is often used to offload the computationally intensive workload to Edge-AI servers \cite{moyukh}, making OHMDs more lightweight. In turn, high-quality video and HMD orientations are transmitted in the UL at first. Then a relatively lower-quality video, strictly synchronized with the UL and with overlaid virtual objects are transmitted in the DL. Transmission of tactile control and haptic feedback are optional, depending on the type of usage scenario. Thus, high-throughput, low-latency, and high-reliability are required for both UL and DL traffic.
    \item[\ding{112}] \textbf{Mixed Reality (MR).} Both the physical and virtual worlds are blended together such that real and active virtual objects coexist and humans can interact with them in real-time. This implies that, unlike in AR, the virtual objects can \emph{occlude}, i.e., be hidden by real-world objects and respond to the geometry, physics, and lighting of the real-world. Hence, the UL and DL data transmission in MR are similar to AR, but more advanced spatial mapping, scene understanding, and latency-sensitive data synchronization are required \cite{mr}.
    \item[\ding{112}] \textbf{Human-to-Human (H2H) collaborations.} Geographically distant human users immersively communicate and collaborate within a shared virtual or mixed environment. Each user’s HMD orientations and tactile control signals are transmitted in the UL, thus demanding low-latency and high-reliability \cite{vr_mpg}. Consequently, high-quality viewpoint-dependent video, reconstructed avatars, and synchronized haptic feedback are transmitted in the DL such that perceptual consistency is maintained across participants.
    \item[\ding{112}] \textbf{Human-to-Machine (H2M) collaborations.} Different to other XR applications, human users interact with virtual or remote machines, such as industrial robots or drones, through tactile and haptic interfaces. Each human user's HMD orientations and tactile control signals are transmitted in the UL to the application server and then forwarded to the remote machine in the DL, as shown in Fig.~\ref{xrc}. Correspondingly, real-time situational video with overlaid environmental data and haptic feedback are transmitted in the UL from the remote machine to the application server and then delivered to the human in the DL. The UL and DL traffic thus form a closed-loop H2M interactive control system relying on high-throughput, low-latency, low-jitter, and high-reliability connections. Nonetheless, the strict latency constraint can be relaxed through decoupling the control-haptic loop using preemptive haptic feedback \cite{h2m_sourav}.
\end{itemize}
%
\par To date, immersive XR applications are increasingly \emph{deployed in domestic and industrial in-premises scenarios, particularly for collaborative design and remote teleoperation}. The success of this demanding ecosystem relies on \emph{three pillars}: advanced software algorithms, efficient hardware devices, and robust network infrastructure. Although XR hardware and software technologies are advancing at a rapid pace, the underlying network often remains the primary bottleneck in delivering an ideal, cyber-sickness-free QoE. Bridging this gap requires moving beyond legacy best-effort indoor setups toward high-capacity, deterministic infrastructures, making advanced coordinated network designs essential.


\begin{figure*}[!t]
\centering
\includegraphics[width=0.8\textwidth,keepaspectratio]{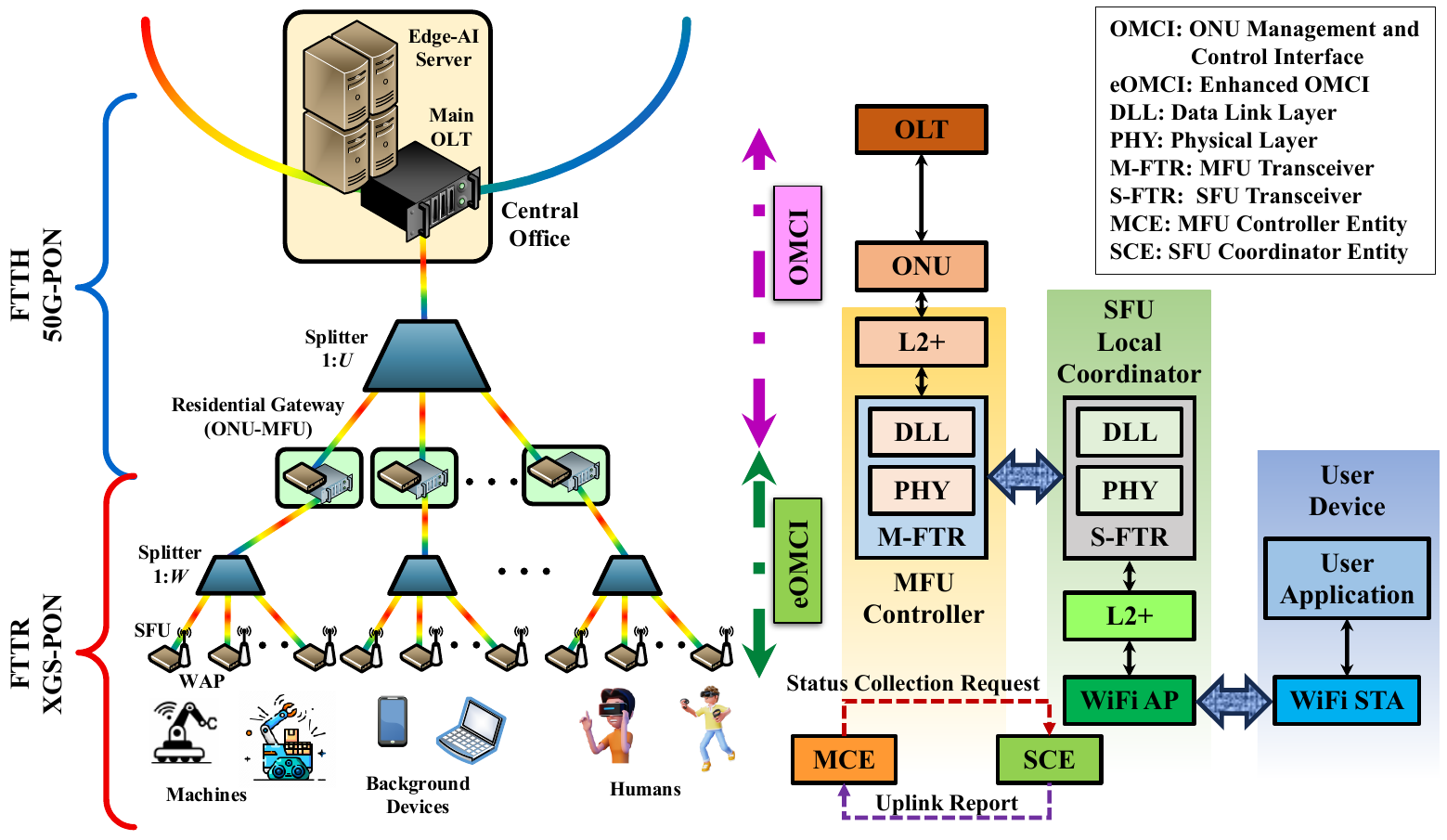}
\caption{A schematic showing the control and management interfaces of the multi-tier integrated FTTR architecture.}
\label{fig3}
\end{figure*}
\setlength{\textfloatsep}{5pt}

\section{Integrated FTTR Network Architecture} \label{sec4}
\subsection{Evolution from FTTH to Integrated FTTR Architecture} \label{sec4.1}
The widespread global success of the FTTH architecture stems from the robust scalability and operational efficiency of G-PON networks, as defined in the ITU-T G.984 series. The initial G-PON deployment in 2004 had nominal DL and UL datarates of 2.5 and 1.25~Gbps. Since then, continuous innovations through XG(S)-PON and NG-PON2 have culminated in the recent standardization of 50G-PON, delivering 50 Gbps per wavelength \cite{50gpon}. At the core of this success lies the OMCI protocol specified in G.988, which introduced a powerful, vendor-agnostic management framework enabling the OLT to centrally configure and control all ONUs. This is achieved through a protocol-independent management information base composed of granular managed entities, allowing operators to remotely provision, configure, and monitor different aspects like the creation of G-PON encapsulation method (GEM) ports for traffic flows and the assignment of transmission containers (T-CONTs) for DBA. However, the traditional G.988 management domain terminates at the residential gateway ONU, leaving in-premise connectivity to low-datarate, best-effort, and high-latency technologies like WiFi or legacy copper. This ``last-meter" connectivity gap is overcome by the integrated FTTR architecture which extends PONs with integrated WAPs in in-premises scenarios, creating a new hierarchical point-to-multi-point (P2MP) topology, as shown in Fig.~\ref{fig3}.
\subsection{Centralized Control and In-Premise Management in FTTR} \label{sec4.2}
Following the external FTTH segment based on 50G-PON, the internal FTTR segment typically comprises an MFU that interfaces with multiple SFUs distributed in individual rooms, interconnected via an in-premise G.fin P2MP (ITU-T G.9940) network architecture \cite{ETSI_GR_F5G032}. For the control and management of this hierarchical FTTR segment, the OMCI is extended to the eOMCI as described in ITU-T G.Sup.eOMCI, where the OLT manages the MFUs in residential gateways as standard ONUs. Moreover, each MFU acts as a proxy OLT for the SFUs, employing an internal OMCI-based management interface for configuration, firmware upgrade, and fault supervision, along with a complementary control interface to support real-time operations \cite{ITUTG9940}. Additionally, the WiFi coordination messages are exchanged via the WLAN management and control interface (WMCI). In this control plane, the MFU controller entity (MCE) sends \emph{status collection requests} to all the SFU coordinator entities (SCEs). In response, the SCEs send \emph{uplink reports} to the MCE, containing airtime utilization, channel conditions, received signal strengths, interference levels, and buffer status reports (BSRs) of all the associated STAs. Using this information, the MCE formulates coordinated strategies for optimal resource allocation and seamless in-premise handovers. The corresponding control commands are then transmitted to the SCEs who execute them locally over both the FTTR and WiFi domains \cite{ITUTG9942}. Note that both the eOMCI and WMCI control messages are multiplexed with user data for transmission as a Type-4 T-CONT (TC-4) flow utilizing the same wavelength(s). To ensure the reliability of this in-band signaling, different IEEE 802.1p Class-of-Service (CoS) priorities can be assigned internally at MFU/SFUs, e.g., the highest priority (CoS-7) for eOMCI and WMCI control messages, followed by HMD orientation, control, and haptics (CoS-6), video (CoS-5), and background traffic (CoS-1).
\begin{figure*}[!t]
\centering
\includegraphics[width=0.9\textwidth,keepaspectratio]{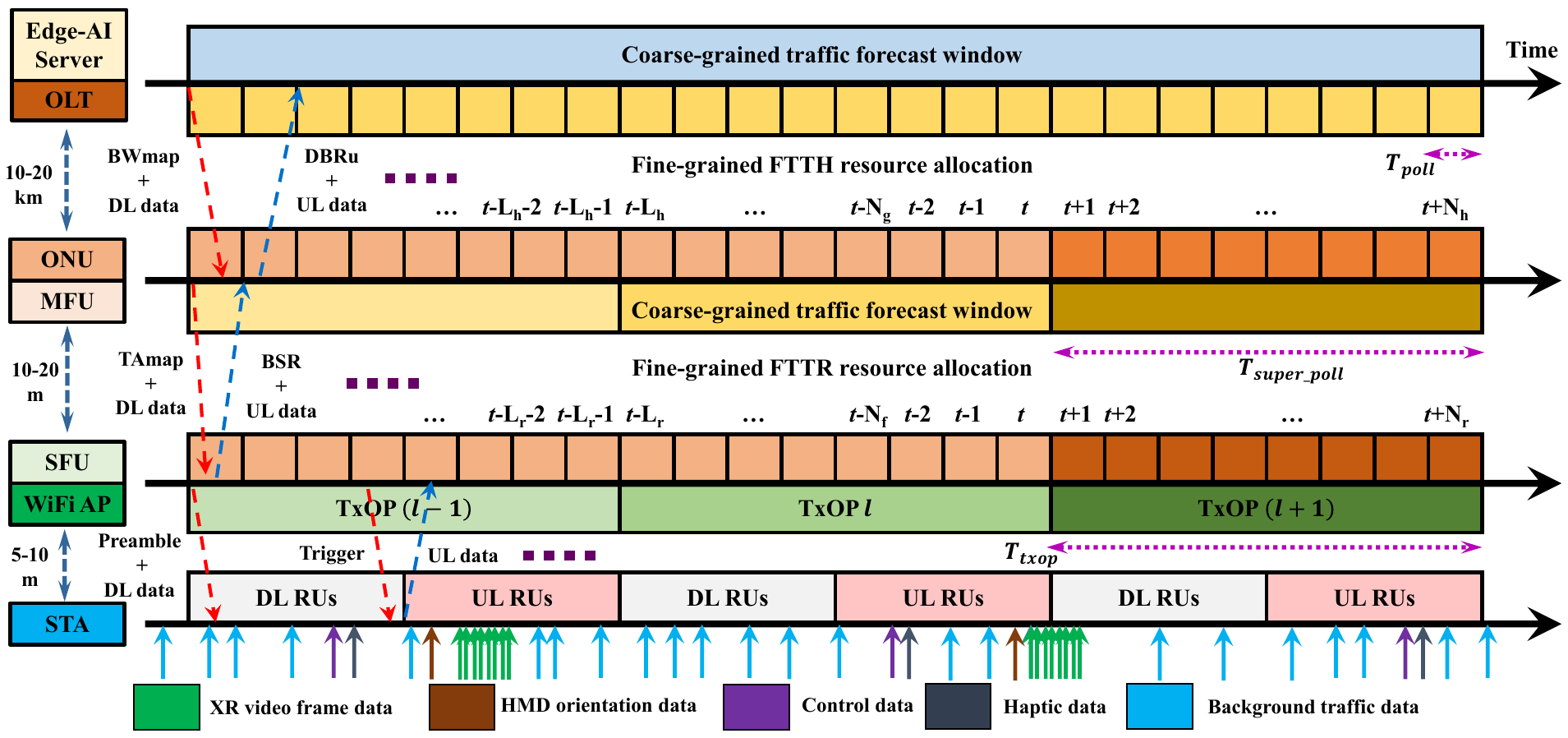}
\caption{Timeline showing the multi-tier coordinated proactive resource allocation in the integrated FTTR network architecture. Time durations of different intervals shown are only for illustration and are not to the scale).}
\label{timing}
\end{figure*}
\setlength{\textfloatsep}{5pt}
\subsection{FTTR-WiFi Integration with Multi-Link Operation} \label{sec4.3}
Since WiFi-6/6E, the adoption of OFDMA has enabled deterministic and contention-free simultaneous channel access for multiple users by dividing the spectrum into orthogonal resource units (RUs) of size $\{2, 4, 8, 20, 40, 80, 160, 320\}$~MHz. Within a \emph{transmission opportunity} (TxOP) duration, firstly the WAP transmits DL data to the associated STAs over their respective RUs and receives acknowledgment from the STAs in the UL. After that, the WAP broadcasts a \emph{trigger frame} containing the RUs allocated to each STA for UL transmissions. Accordingly, the STAs transmit their UL data and receive a block acknowledgment frame in the DL \cite{wifi7}. Usually in in-premise scenarios, the densely deployed WAPs are configured with a distinct channel of size $[20, 40, 80, 80+80, 160, 320]$~MHz in the 5~GHz or 6~GHz band. Although the choice of bigger channels make spectrum reuse very frequent, the MFUs can centrally create contention-free transmission times of the STAs, intimated by trigger messages and associated to neighboring WAPs operating with the same channels. Nevertheless, the MLO introduced since WiFi-7 enables the WAPs to simultaneously operate in $[2.4, 5, 6]$~GHz bands with different bandwidths. For instance, in an XR collaboration, high-datarate video traffic can be assigned to the 6~GHz link to leverage its 320~MHz channels, while HMD orientation, control, haptic, and background traffic can be simultaneously assigned to a 160~MHz channel in 5~GHz band utilizing smaller 20-40~MHz RUs.
\subsection{System Model and Coordinated Resource Allocation} \label{sec4.4}
In this work, we consider the internal FTTR segment consists of a XG(S)-PON supporting WiFi-7 WAPs. Each WAP has 16 antennas, equally divided for simultaneous operations in both 5~GHz and 6~GHz bands. Moreover, all WAPs in each premise are configured with the same TxOP duration $T_{txop}$ (maximum values for voice and video access classes are 1.505 and 3.008 msec, respectively), much bigger than the $T_{poll}$ = 125~$\mu$sec polling cycle of XGS-PON. To design a timeslotted FTTR-WiFi coordinated resource allocation scheme, as shown in Fig.~\ref{timing}, we consider $T_{txop}$ as \emph{an integer multiple} of $T_{poll}$, e.g., 1-3 msec. Note that the OMCI-based interfaces have inherent mechanisms for time synchronization among all the nodes. After an MFU receives BSRs and uplink reports from the associated SFU-WAPs in a TxOP, it schedules the RUs and transmission times of the WAPs in the DL and the STAs in the UL for the next TxOP. In addition, the corresponding TAmaps for the intermediate $\lceil T_{txop}/T_{poll} \rceil$ polling cycles are also generated. Note that the G.fin TAmap is functionally analogous to the G-PON BWmap, but its duration needs not to be fixed at 125 $\mu$sec \cite{ITUTG9942}. It can operate on a longer, configurable \emph{super-frame}, i.e., $T_{super\_poll}$ = 1-3 msec to align with WiFi timeslots. The complex aggregated multi-modal XR and background traffic at the residential gateways are reported further in each polling cycle to the OLT of the external FTTH segment based on 50G-PON. Accordingly, BWmaps are generated for transmission of this data in the UL. Since the FTTH, FTTR, and WiFi modules largely operate independently in practical scenarios, we adopt this \emph{coarse-grained resource allocation scheme} for its efficiency and ease of rapid deployment. Nonetheless, its reactive nature hinders meeting the stringent QoE requirements of XR collaborations. Hence, we design a proactive resource allocation framework empowered by state-of-the-art neural network and Transformer-based long-horizon multi-step-ahead traffic forecasting methods.\par

\section{Coarse-Grained Resource Allocation} \label{sec5}
In this section, we propose the framework embedding highly-accurate time-series forecasting methods to enable coarse-grained proactive resource allocation across the internal FTTR and external FTTH segments. In addition, seamless in-premise intelligent roaming schemes are devised to satisfy the stringent QoE requirements of XR collaborations.\par
%
\subsection{Traffic Forecasting in FTTR Segment} \label{sec5.1}
Recall that $T_{txop}$ denotes the TxOP duration configured for all the WAPs in a premise and $T_{poll}$ denotes the polling cycle duration. Therefore, within each TxOP, there are $N_r = \lceil T_{txop}/T_{poll}\rceil$ XG(S)-PON polling cycles or timeslots. Let $\mathcal{I} = \{1,\dots,W\}$ denote the set of SFU/WAPs, $\mathcal{J} = \{1,\dots,S_{xr},S_{xr}+1,\dots,S_{xr}+S_{bkg}\}$ denote the set of STAs connected to each WAP, where $S_{xr}$ indicates the number of XR devices with either video, HMD orientation, tactile control, or haptic traffic and $S_{bkg}$ indicates the number of background devices, $x_t^{(i,j)}$ denote the UL traffic from STA-$j$ to WAP-$i$ in timeslot-$t$ and $X_l^{(i,j)} = \sum_{k=1}^{N_r} x_{t+k}^{(i,j)}$ denotes the total traffic arrival in TxOP-$l$ enveloping $N_r$ timeslots. We represent the input sequence of aggregated UL traffic at WAP-$i$ over the last $L_r$ (usually $L_r\gg N_r$) timeslots as,\\
\vspace{-\baselineskip}
\begin{align}
 \bm{X}_{t-1}^{(i)} &= [\bm{x}_{t-L_r}^{(i)}, \dots, \bm{x}_{t-1}^{(i)}], \forall i\in\mathcal{I}, \label{eq1}\\
 \bm{x}_{t-1}^{(i)} &= [x_{t-1}^{(i,j)}]_{j=1}^{S} =[x_{t-1}^{(i,1)}, \dots, x_{t-1}^{(i,S)}]^\top, \forall i\in\mathcal{I}. \label{eq2}
\end{align}
\par Considering that traffic characteristics of each STA will be independent of each other, a separate long-horizon multi-step-ahead traffic forecast model $f_{\Theta}^{(i,j)}(\cdot)$ for each STA is created at the MFU. Nonetheless, common forecast models can be created for STAs with similar traffic and their historical data can be combined for model training. If the model inference latency $\tau_r \geq T_{poll}$, then the inferencing must be initiated $N_f = \lceil{\tau_r/T_{poll}}\rceil$ timeslots before the current TxOP ends. We adopt the 95$^{\text{th}}$ percentile of the inference time distribution as $\tau_r$. Let the inference completion time be denoted as timeslot-$t$. Consequently, the inference process was initiated at timeslot $(t-N_f)$. Using an input sequence ending at $(t-N_f)$, denoted as $\mathbf{x}_{t-N_f}^{(i,j)} = [x_{t-N_f-L_r+1}^{(i,j)}, \dots, x_{t-N_f}^{(i,j)}]^\top$, the model generates forecasts for the next $(N_f+N_r)$ timeslots:\\
\vspace{-\baselineskip}
\begin{align}
\hat{\bm{y}}_{t-N_f}^{(i,j)} &= f_{\Theta}^{(i,j)}(\mathbf{x}^{(i,j)}_{t-N_f}), \nonumber\\
&= [\hat{x}_{t-N_f+1}^{(i,j)}, \dots, \hat{x}_t^{(i,j)}, \hat{x}_{t+1}^{(i,j)}, \dots, \hat{x}_{t+N_r}^{(i,j)}]^\top. \label{{eq3}}
\end{align}
Since $N_f$ timeslots are consumed to make the traffic forecasts, the predictions $[\hat{x}_{t-N_f+1}^{(i,j)}, \dots, \hat{x}_t^{(i,j)}]$ correspond to the past and are discarded. Only the sequence $[\hat{x}_{t+1}^{(i,j)}, \dots, \hat{x}_{t+N_r}^{(i,j)}]^\top$ is utilized to generate TAmaps for the next $N_r$ polling cycles. In each polling cycle or timeslot-$t$, after receiving BSR $r_t^{(i,j)}$ and data $\alpha_t^{(i,j)}$, the MFU calculates the actual traffic arrival for the previous slot as $x_{t-1}^{(i,j)} = \max\{0, r_t^{(i,j)}+\alpha_t^{(i,j)}-r_{t-1}^{(i,j)}\}$. This relation holds
true for any PON where the one-way transmission latency
is less than $T_{poll}$. For the immediate next timeslot-$(t+1)$, the MFU grants each SFU-$i$ with:
$g_{t+1}^{(i,j)} = \min\left\{\rho_{t+1}^{(i,j)}, \left(\frac{\rho_{t+1}^{(i,j)}}{\sum_k \rho_{t+1}^{(i,k)}}\right) \cdot BW_{max}^{(i)}\right\}$,
where $\rho_{t+1}^{(i,j)} = (r_t^{(i,j)} + \hat{x}^{(i,j)}_{t+1})$, representing the \emph{projected buffer occupancy}, i.e., the sum of the currently reported buffer status ($r_t$) and the new traffic predicted to arrive during the grant execution ($\hat{x}_{t+1}$), and $BW_{max}^{(i)} = (R_r \times T_{poll}/W)$ denotes the maximum bandwidth available per polling cycle where $R_r$ denotes the line rate of the PON in FTTR segment. It is interesting to note that this resource allocation scheme is robust against under and over-predictions. To train this model with a historical dataset, the mean-squared error (MSE) loss across the $(N_f+N_r)$ timeslots per-sample-$\mu$ is given by,\\
\vspace{-\baselineskip}
\begin{align}
\mathcal{L}_{\mu}^{(i,j)} = \frac{1}{N_f+N_r}\sum_{k=-N_f+1}^{N_r} \left(x_{t+k}^{(i,j)} - \hat{x}_{t+k}^{(i,j)}\right)^2. \label{eq4}
\end{align}

\par For detailed discussion of the models, see Appendix A.


\subsection{Traffic Forecasting in FTTH Segment} \label{sec5.2}
All traffic generated within each FTTR-WiFi segment are expected to be aggregated at the corresponding ONU of the FTTH segment and yields a highly complex traffic pattern. Consequently, traditional resource allocation schemes in external FTTH networks face severe operational bottlenecks in satisfying the QoE requirements of XR collaborations. Recall that we consider that all traffic belong to TC-4 flows and we may proactively allocate resources only for multi-modal XR traffic in both FTTH and FTTR-WiFi segments as background traffic is low-priority and latency-tolerant. Let $\xi_t^{(u)}$ and $\beta_t^{(u)}$ denote the aggregated multi-modal XR traffic and background traffic arrival at ONU-$u$ in polling cycle or timeslot-$t$, respectively. The coarse-grained interval for traffic forecasting $T_{int}$ is chosen to be a large-integer multiple of polling cycle duration such that $N_h = \lceil T_{int}/T_{poll}\rceil$, as shown in Fig.~\ref{timing}. If the traffic forecast model inference latency (95$^{\text{th}}$ percentile) $\tau_h \geq T_{poll}$, then the inferencing should be started $N_g = \lceil{\tau_h/T_{poll}}\rceil$ timeslots before the target interval $T_{int}$ begins. This requires forecasting for an additional $N_g$ timeslots to cover the inference latency. This expanded inference window inherently requires the adoption of an advanced Transformer-based model, which significantly outperforms conventional recurrent and convolutional neural networks in long-horizon, multi-sample-ahead forecasting \cite{bishop2023deep}.\par

The sequence of aggregated multi-modal XR traffic from all the $U$ ONUs over the last $L_h$ (usually $L_h\gg N_h$) polling cycles, available at the start of inference, is represented as:\\
\vspace{-\baselineskip}
\begin{align}
\mathbf{\Xi}_{t-N_g} &= [\bm{\xi}_{t-N_g-L_h+1}, \ldots, \bm{\xi}_{t-N_g}], \label{eqf1}\\
\bm{\xi}_{k} &= [\xi_{k}^{(1)}, \ldots, \xi_{k}^{(U)}]^\top. \label{eqf2}
\end{align}
Using \eqref{eqf1} as input data, for each ONU-$u$ the Transformer-based traffic forecasting module implemented in the Edge-AI server generates $\hat{\bm{\Xi}}^{(u)}_{t-N_g+1:t+N_h} = \gamma^{(u)}_\Phi(\mathbf{\Xi}^{(u)}_{t-N_g})$, where $\gamma^{(u)}_{\Phi}(.)$ is a parameterized multi-step predictor implemented using \emph{attention mechanisms}. The complete forecast generated is $\hat{\bm{\Xi}}_{t-N_g+1:t+N_h} = [\hat{\bm{\xi}}_{t-N_g+1}, \dots, \hat{\bm{\xi}}_{t}, \hat{\bm{\xi}}_{t+1}, \dots, \hat{\bm{\xi}}_{t+N_h}]$. Upon completion of inference which started at timeslot-$(t-N_g)$ and finished at timeslot-$t$, the first $N_g$ predictions $[\hat{\bm{\xi}}_{t-N_g+1}, \dots, \hat{\bm{\xi}}_{t}]$ are discarded as these timeslots have already passed. The remaining traffic forecasts $\hat{\bm{\xi}}^{(u)}_{t+k}, \forall k\in\{1,\dots,{N_h}\}$ for each ONU are used by the BWmap generator to proactively allocate resources for uplink transmission.

We consider that the maximum allocable bandwidth to each ONU-$u$ in a polling cycle is $BW^{(u)}_{max} = (R_h\times T_{poll}/U)$, where $R_h$ denotes the line rate of the PON in the FTTH segment. This bandwidth is proportionally divided between the XR and background traffic to minimize bandwidth wastage. Therefore, after receiving a dynamic buffer report (DBRu) containing $rx_t^{(u)}$ for XR traffic and $rb_t^{(u)}$ for background traffic in timeslot-$t$, the corresponding grant provided to ONU-$u$ consists of $g{\xi}_{t+1}^{(u)} = \min\left\{ \rho \xi_t^{(u)},\frac{\rho \xi_t^{(u)}}{\rho\xi_t^{(u)}+rb_t^{(u)}}\cdot BW^{(u)}_{max}\right\}$ and $g{\beta}_{t+1}^{(u)} = \min\left\{rb_t^{(u)},\frac{rb_t^{(u)}}{\rho\xi_t^{(u)}+rb_t^{(u)}}\cdot BW^{(u)}_{max}\right\}$, where $\rho\xi_t^{(u)}= (rx_t^{(u)}+\hat{\xi}_{t+1}^{(u)})$. Note that $\hat{\xi}_{t+1}^{(u)}$ is used to account for arrivals during the grant execution. After receiving XR data $\alpha x_t^{(u)}$, the actual XR traffic arrival in the previous timeslot-$(t-1)$ is calculated as $\xi_{t-1}^{(u)} = \max\{0, rx_t^{(u)}+\alpha x_t^{(u)}-rx_{t-1}^{(u)}\}$. The training process of the traffic forecast models for each ONU-$u$ aims to minimize the multi-step MSE loss across the $(N_g+N_h)$ timeslots per-sample-$\nu$, given by:\\
\vspace{-\baselineskip}
\begin{align}
\mathcal{L}_{\nu}^{u} = \frac{1}{N_g+N_h}\sum_{k=-N_g+1}^{N_h}
\left(\xi_{t+k}^{(u)} - \hat{\xi}_{t+k}^{(u)}\right)^2. \label{eqf3}
\end{align}
\par Detailed descriptions of the models are provided in Appendix B.

\subsection{Intelligent In-Premise Seamless Roaming} \label{sec5.3}
We formulate a mixed-integer non-linear program (MINLP), denoted as $\mathcal{P}_{FTTR}$ for an intelligent in-premise optimal WAP-STA association, resource allocation, and seamless handover. The primary objective is to \emph{maximize the total network utility in each TxOP}, ensuring that the UL and DL resource demands of a maximum number of STAs are satisfied. Initially, all STAs report the channel conditions of their neighboring WAPs to be assigned to a WAP with good channel condition and sufficient capacity to meet its demands. To ensure stability, associated STAs remain attached to their current WAPs unless their QoE requirements, especially throughput, latency, and reliability, are violated over $N_{\delta}$ consecutive TxOPs. Upon such violation, the MFU initiates a ``make-before-break" handover for the STAs experiencing poor channel conditions as described in \cite{fttrb_ofc}. Immediately these STAs are requested to report the channel conditions of other neighboring WAPs. After receiving their uplink reports, in the subsequent TxOP, the MFU solves $\mathcal{P}_{FTTR}$ to identify the optimal target WAP and transmits a handover indication to the OLT via the FTTH segment to reconfigure the network path. Once the network paths are reconfigured with new WAPs, the corresponding STAs are handed over to the new WAPs seamlessly (refer to Fig. \ref{arch}). Practically, this scheme minimizes frequent handovers of mobile STAs within the dense in-premise FTTR-WiFi segment.\par
\textbf{Sets and Parameters:} Recall that $\mathcal{I} = \{1,\dots,W\}$ denotes the set of WAPs, $\mathcal{J} = \{1,\dots,S_{xr},S_{xr}+1,\dots,S_{xr}+S_{bkg}\}$ denote the set of STAs, and $X_l^{(i,j)}$ represents the UL traffic arriving at WAP-$i$ from STA-$j$ during OFDMA TxOP-$l$. The set of MLO frequency bands supported by the devices is $\mathcal{B} = \{2.4, 5, 6\}$ GHz and the corresponding sets of available RUs with bandwidth in MHz are $\mathcal{W}_{2.4} = \{2,4,8,20,40\}$, $\mathcal{W}_{5} = \{2,4,8,20,40,80,160\}$, and $\mathcal{W}_{6} = \{2,4,8,20,40,80,160,320\}$. Let $BW_b^{max}$ denote the total bandwidth capacity of band $b \in \mathcal{B}$. Channel quality is characterized by the spectral efficiency $E_{ijb}$ (bits/s/Hz) for the link between WAP-$i$ and STA-$j$ on band $b$, weighted by a priority factor $p_{ijb}$. We define $R_{req,j}^{UL}$ and $R_{req,j}^{DL}$ as the minimum data rate per TxOP for STA-$j$ (see Table \ref{table1}).To enforce seamless roaming with hysteresis, we utilize the parameter $\theta_{ij} \in \{0,1\}$, which indicates the association status in the previous TxOP. Assuming traffic forecasting and resource allocation decision-making require $N_w$ TxOPs, the forecast horizon is set to $(N_w+1)$. At the end of TxOP-$l$, upon receiving the BSR $r_l^{(i,j)}$ and data $X_l^{(i,j)}$, the MFU calculates the traffic generated in the previous TxOP as $\Gamma_{l-1}^{(i,j)} = \max\{0, r_l^{(i,j)}+X_l^{(i,j)}-r_{l-1}^{(i,j)}\}$. Using an input sequence of length $L_w$, denoted as $\mathbf{\Gamma}_{l-N_w}^{(i,j)} = [\Gamma_{l-N_w-L_w+1}^{(i,j)}, \dots, \Gamma_{l-N_w}^{(i,j)}]^\top$, the MFU generates forecasts $\hat{\bm{\Gamma}}_{l-N_w+1:1}^{(i,j)}=[\hat{\Gamma}^{(i,j)}_{l-N_w+1},\dots,\hat{\Gamma}^{(i,j)}_{l},\hat{\Gamma}^{(i,j)}_{l+1}]$ for the next $(N_w+1)$ TxOPs via the predictive model detailed in Sec. \ref{sec5.1}. As $N_w$ TxOPs elapse during inference, we utilize $\hat{\Gamma}^{(i,j)}_{l+1}$ to project the UL buffer occupancy of STA-$j$ as $Q_{i,j}^{UL} (l+1) = (r_l^{(i,j)}+\hat{\Gamma}^{(i,j)}_{l+1})$. The DL buffer occupancy $Q_{i,j}^{DL} (l+1)$ is deterministically known by the MFU. These values, $Q_{i,j}^{DL}$ and $Q_{i,j}^{UL}$, represent the \emph{resource demands} of STA-$j$ and are transmitted to WAP-$i$ immediately.\par
%
\textbf{Decision variables:} Let $\phi_{ij} \in \{0,1\}$ be a binary variable where $\phi_{ij}=1$ if STA-$j$ associates with WAP-$i$, and 0 otherwise. For resource allocation, we introduce binary variables $\mu_{ijbw}^{DL}, \mu_{ijbw}^{UL} \in \{0,1\}$, which equal to 1 if STA-$j$ is assigned a discrete RU of bandwidth $w \in \mathcal{W}_b$ on band $b$ of WAP-$i$. Additionally, $\tau_i \in [\tau_{min}, \tau_{max}]$ denotes the continuous time fraction of the total TxOP duration $T_{txop}$ allocated to the downlink transmission for WAP-$i$, with a guard time $\tau_g$ maintained between DL and UL intervals.\par
\textbf{MINLP formulation:} The \emph{objective function} of $\mathcal{P}_{FTTR}$ maximizes the total utility of satisfied demands:\\
\vspace{-\baselineskip}
\begin{align}
\max_{\bm{\phi},\bm{\mu},\bm{\tau}} \sum_{i \in \mathcal{I}} \sum_{j \in \mathcal{J}} \phi_{ij} \Bigg[ & \sum_{b\in\mathcal{B}} p_{ijb} \Biggl( \min\left\{Q_{ij}^{DL}, R_{ij}^{DL}\right\} \Biggr. \nonumber\\
& \quad \Biggl. + \min\left\{Q_{ij}^{UL}, R_{ij}^{UL} \right\}\Biggr) \Bigg], \label{obj}
\end{align}
where the $\min\{\cdot\}$ function ensures the utility is capped by the buffer occupancy if capacity $R_{ij}$ exceeds demand $Q_{ij}$. Conversely, if $R_{ij} < Q_{ij}$, the utility is limited by the feasible data transmission, implying excess data is queued for the next TxOP and increasing latency in turn. The feasible DL and UL data transmissions $\forall i\in\mathcal{I}, j\in \mathcal{J},$ are defined as:\\
\vspace{-\baselineskip}
\begin{align}
R_{ij}^{DL} &= \tau_i \cdot T_{txop} \cdot \sum_{b\in\mathcal{B}} \bigg(E^{DL}_{ijb} \cdot \sum_{w\in\mathcal{W}_b} \mu^{DL}_{ijbw} \cdot w \bigg), \label{eqR1}\\
R_{ij}^{UL} &= (1-\tau_i-\tau_g) \cdot T_{txop} \cdot \sum_{b\in\mathcal{B}} \bigg(E_{ijb}^{UL} \cdot \sum_{w\in\mathcal{W}_b} \mu^{UL}_{ijbw} \cdot w \bigg). \label{eqR2}
\end{align}
In \eqref{eqR1}-\eqref{eqR2}, the spectral efficiency terms are given by:\\
\vspace{-\baselineskip}
\begin{align}
E^{DL}_{ijb} &= \log_2 \Bigg(1+\frac{P_{Tx}^{i}\cdot G^{DL}_{ijb}}{N_0 + I_{ijb}^{DL}}\Bigg), \quad \forall i, j, b,\label{eqE1} \\
E^{UL}_{ijb} &= \log_2 \Bigg(1+\frac{P_{Tx}^{j}\cdot G^{UL}_{ijb}}{N_0 + I_{ijb}^{UL}}\Bigg), \quad \forall i, j, b, \label{eqE2}
\end{align}
where $P_{Tx}^{i}$ and $P_{Tx}^{j}$ denote the transmit power of the WAP and STA, respectively, $G_{ijb}$ denotes the channel gain (including path loss and fading), $I_{ijb}$ represents the measured interference, and $N_0$ is the noise power.\par
\emph{Throughput Constraints:} If a STA is connected to a WAP, i.e., $\phi_{i,j} = 1$, the aggregated throughput per TxOP must meet the minimum QoE requirements:\\
\vspace{-\baselineskip}
\begin{align}
R_{ij}^{DL} &\ge \phi_{ij} \cdot R_{req, j}^{DL}, \quad \forall i,j, \label{eqass1}\\
R_{ij}^{UL} &\ge \phi_{ij} \cdot R_{req, j}^{UL}, \quad \forall i,j. \label{eqass2}
\end{align}
These inequalities ensure that if STA-$j$ is not associated with WAP-$i$ ($\phi_{ij}=0$), the required rate from that specific WAP is zero, satisfying the constraint trivially.\par
\emph{Connectivity constraints:} Each STA may associate with at most one WAP. To minimize ping-pong handovers, existing connections are maintained unless QoE is violated. The parameter $\theta_{ij}$ captures the previous association state. It is initialized to 0 and updated to 1 once a connection is established. If QoE degradation persists for some STA-$j$ in consecutive $N_{\delta}$ TxOPs, then these constraints are relaxed to find a new WAP and network-initiated handover is performed over the next TxOPs.\\
\vspace{-\baselineskip}
\begin{align}
    \sum_{i \in \mathcal{I}} \phi_{ij} &\le 1, \quad \forall j, \label{eqass3}\\
    \phi_{ij} &\ge \theta_{ij}, \quad \forall i, j. \label{eq:hysteresis}
\end{align}
\begin{algorithm}[t!]
\caption{FTTR-WiFi Resource Allocation and Roaming} \label{alg1}
\hspace*{\algorithmicindent} \textbf{Input:} $\mathcal{I}, \mathcal{J}, \mathcal{B}, \mathcal{W}_b, Q_{ij}^{DL/UL}, E_{ijb}^{DL/UL}, \theta_{ij}, p_{ijb,} N_{max}, \epsilon$\\
\hspace*{\algorithmicindent} \textbf{Output:} Optimal solution: $\bm{\phi}^*, \bm{\mu}^*, \bm{\tau}^*$\\
\hspace*{\algorithmicindent} \textbf{Initialize:} $n \leftarrow 0$, $\tau_i^{(0)} \leftarrow 0.5, \ \forall i$, Objective $U^{(0)} \leftarrow 0$;
\begin{algorithmic}[1]
\While{$n < N_{max}$}
    \State \texttt{\# STA Association and RU Allocation}
    \State Fix time allocation $\bm{\tau}^{(n)}$;
    \For{$i \leftarrow 1$ \textbf{to} $|\mathcal{I}|$} \Comment{\textit{Parallelizable across WAPs}}
        \State Update $R_{ij}^{DL/UL}$ capacities based on fixed $\tau_i^{(n)}$ using (\ref{eqR1})-(\ref{eqR2});
        \State Employ greedy matching to find $\bm{\phi}_i^{(n+1)}, \bm{\mu}_i^{(n+1)}$;
        \State Ensure constraints (\ref{eqass1})-(\ref{eqcap3}) are satisfied;
    \EndFor
    \State \texttt{\# Time Allocation Optimization}
    \State Fix association $\bm{\phi}^{(n+1)}$ and RU allocation $\bm{\mu}^{(n+1)}$;
    \For{$i \leftarrow 1$ \textbf{to} $|\mathcal{I}|$}
        \State Calculate optimal $\tau_i^{(n+1)}$ via gradient update;
        \State Clip $\tau_i^{(n+1)}$ to $[\tau_{min}, \tau_{max}]$ to satisfy (\ref{eqtime1});
    \EndFor
    \State \texttt{\# Convergence Check}
    \State Calculate current total objective $U^{(n+1)}$ using (\ref{obj});
    \If{$|U^{(n+1)} - U^{(n)}| \le \epsilon$}
        \State \textbf{break}; \Comment{\textit{convergence or fixed point}}
    \Else
        \State $n \leftarrow n + 1$;
    \EndIf
\EndWhile
\State \textbf{return} $\bm{\phi}^{(n+1)}, \bm{\mu}^{(n+1)}, \bm{\tau}^{(n+1)}$;
\end{algorithmic}
\end{algorithm}
\setlength{\textfloatsep}{5pt}
%
\emph{OFDMA-MLO RU allocation constraints:} These ensure that each STA is assigned at most one discrete RU per band, and the cumulative allocated bandwidth does not exceed $BW_b^{max}$:\\
\vspace{-\baselineskip}
\begin{align}
\sum_{w \in \mathcal{W}_b} \mu^{DL}_{ijbw} \le \phi_{ij} \quad \text{and} \quad &\sum_{w \in \mathcal{W}_b} \mu^{UL}_{ijbw} \le \phi_{ij}, \forall i,j,b, \label{eqcap1}\\
\sum_{j\in\mathcal{J}} \sum_{w \in \mathcal{W}_b} \mu^{DL}_{ijbw} \cdot w &\le BW_b^{max}, \quad \forall i,\label{eqcap2}\\
\sum_{j\in\mathcal{J}} \sum_{w \in \mathcal{W}_b} \mu^{UL}_{ijbw} \cdot w &\le BW_b^{max}, \quad \forall i.\label{eqcap3}
\end{align}
\par \emph{Time allocation constraints:} To ensure valid time-splits and prevent resource starvation for control overhead, e.g., beacons, $\tau_i$ is bounded by guard times \cite{wifi6_ofdma} and the total allocated time, including the guard interval $\tau_g$, is normalized to the TxOP duration such that:\\
\vspace{-\baselineskip}
\begin{align}
\tau_{min} \le \tau_i &\le \tau_{max}, \quad \forall i, \label{eqtime1}
\end{align}
where the constraint $\tau_i + (1-\tau_i-\tau_g) + \tau_g = 1$ is implicitly satisfied by the feasible data transmission definitions \eqref{eqR1}-\eqref{eqR2}.\par
\textbf{Solution algorithm:} Since $\mathcal{P}_{FTTR}$ is equivalent to the \emph{generalized assignment problem}, a well-known NP-hard problem, and contains coupled binary variables ($\phi_{ij}, \mu_{ijbw}$) and continuous variables ($\tau_i$), a \emph{decomposition-based alternating optimization approach} is chosen due to its robustness and time-efficiency \cite{LiSun_Chapter13}. This method separates the problem into two tractable sub-problems and solves them iteratively, as summarized in Alg.~\ref{alg1}. The computational complexity of the proposed Alg.~\ref{alg1} is determined by the number of outer iterations and the cost of solving the two sub-problems. The user association and RU allocation phase, if solved via a greedy heuristic like \cite{minlp}, has a worst-case complexity of $\mathcal{O}(|\mathcal{I}| \cdot |\mathcal{B}| \cdot \max_b\{|\mathcal{W}_b|\} \cdot |\mathcal{J}|)$, where sorting demands and assigning RUs dominates the cost. The time allocation phase is a linear optimization over $|\mathcal{I}|$ variables, solvable in $\mathcal{O}(|\mathcal{I}|^{3.5})$ using interior point methods, or $\mathcal{O}(|\mathcal{I}|)$ using simple gradient updates. Thus, the total asymptotic time complexity is $\mathcal{O}\left(N_{iter} \cdot (|\mathcal{I}| \cdot |\mathcal{J}| \cdot |\mathcal{B}| \cdot \max_b\{|\mathcal{W}_b|\} + |\mathcal{I}|^{3.5})\right)$, where $N_{iter}$ denotes the number of iterations required for convergence. Since $|\mathcal{I}|$, $|\mathcal{J}|$, and $\max_b\{|\mathcal{W}_b|\}$ are finite and small in in-premise FTTR-WiFi scenarios, the algorithm is polynomial-time and feasible for practical deployment.\par

\section{Performance Evaluation} \label{sec6}
In this section, we evaluate the performance of the proposed proactive resource allocation and intelligent in-premise handover framework for the integrated FTTR network architecture adopting neural network and Transformer-based traffic forecast models.\par

\subsection{Network Simulation Configurations} \label{sec6.1}
\begin{figure}[!t]
\centering
\includegraphics[width=\columnwidth,keepaspectratio]{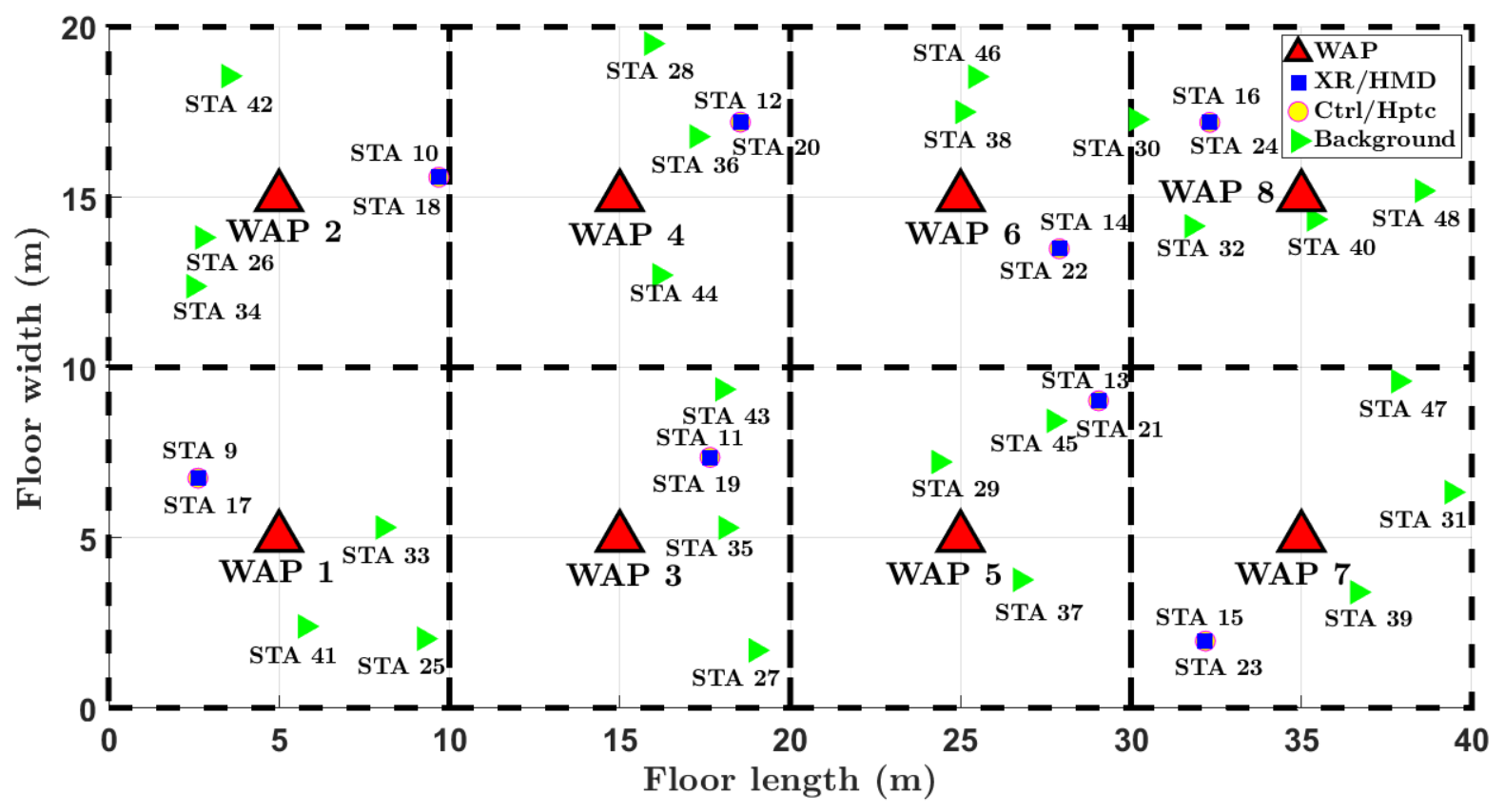}
\caption{Considered FTTR-WiFi scenario with 8 WAPs each associated to 5 STAs supporting XR collaborations and background traffic.}
\label{exp_scn}
\end{figure}

We implement the resource allocation schemes of all the segments of the integrated FTTR architecture using the discrete event-driven simulator OMNeT++. Following the ITU-T recommendations \cite{itu_fttr}, we consider a 50G-PON of 20 km length as the first-segment FTTH network with the maximum available datarate in both the UL and DL are 50 Gbps. The duration of the polling cycles is fixed at 125~$\mu$sec and the guard time between GEM packets from consecutive ONUs is 0.25~$\mu$sec. The number of ONUs controlled by the OLT of the 50G-PON is 32 (split-ratio 1:32). Each of these ONUs are connected to an MFU which control a FTTR segment with XGS-EPON with maximum UL and DL datarate 10~Gbps, split ratio 1:8, MFU-SFU distance 20 m, polling cycle duration 125~$\mu$sec, and the guard time between packets from consecutive SFUs is 0.25~$\mu$sec. Hence, there are a total of 32 × 8 = 256 SFUs. Each SFU is further connected to a WiFi-7 WAP configured with OFDMA fixed TxOP duration of 1 msec and MLO with 8 antennas on 5 GHz and 8 antennas on 6 GHz bands with transmit 20 dBm power in the DL, as shown in Fig. \ref{exp_scn}. Moreover, we consider that 4 human-machine pairs in each premise are participating in an XR collaboration. In each room of dimension 10$\times$10 meter$^2$, the blue square indicates either a human (UL: HMD orientations, DL: XR video) or a machine (UL: XR video, DL: HMD orientations). Correspondingly, the yellow circle beneath it indicates a control device/gloves for humans (UL: control, DL: haptics) or a haptic device (UL: haptics, DL: control). In addition, there are 3 background devices (mobiles, laptops etc.) transmitting exponentially distributed traffic both in the UL and DL. Thus, there are 5 wireless devices or STAs in each room and the locations of all the 8$\times$5 = 40 STAs are randomly generated. Their channel gains are calculated using ray-tracing methods in MATLAB. We configure the XR video devices with 2 antennas and UL transmit power 18 dBm in the 6 GHz band. All other devices are configured with 2 antennas and UL transmit power 18 dBm in the 5 GHz band.\par
\setlength{\textfloatsep}{5pt}
\begin{figure*}[!t]
  \centering
  \subfloat[XR video]{%
    \includegraphics[width=0.333\textwidth]{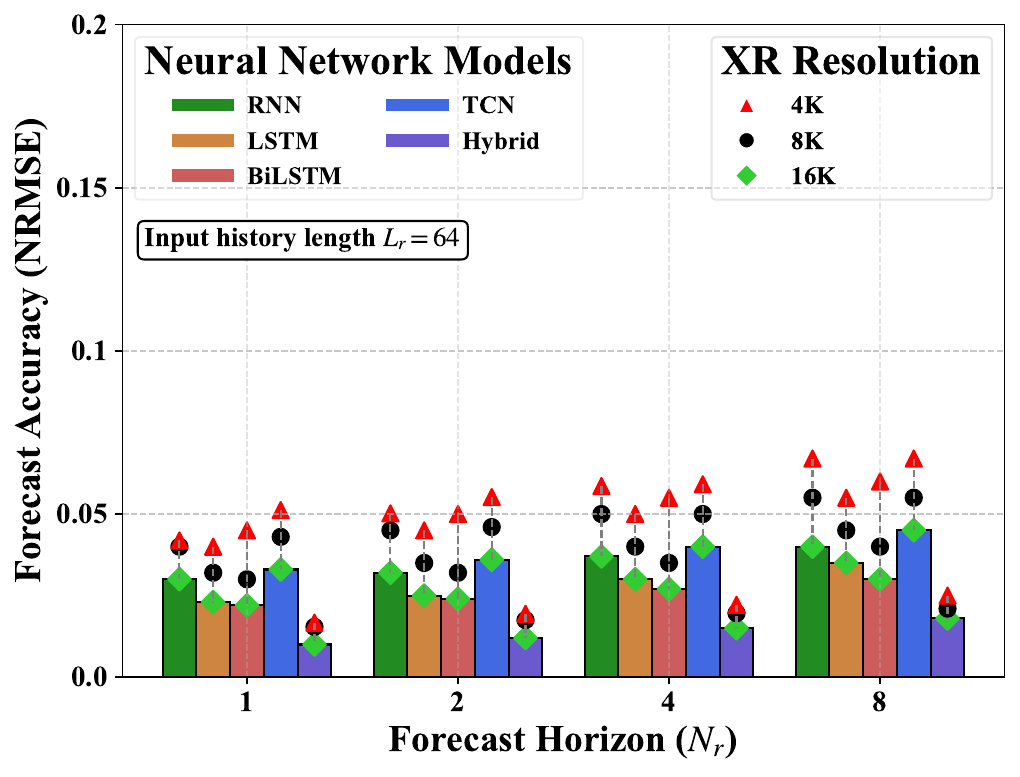}\label{rmsse11}%
  }
  \subfloat[HMD orientations]{%
    \includegraphics[width=0.333\textwidth]{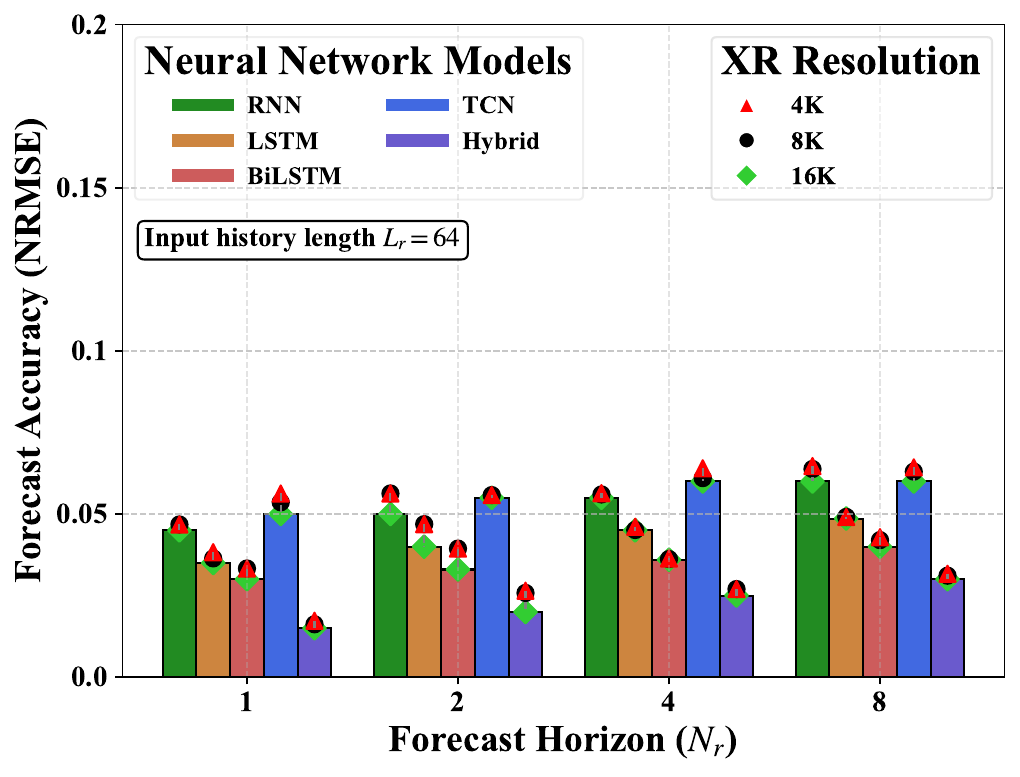}\label{rmsse12}%
  }
  \subfloat[Control and Haptics]{%
    \includegraphics[width=0.333\textwidth]{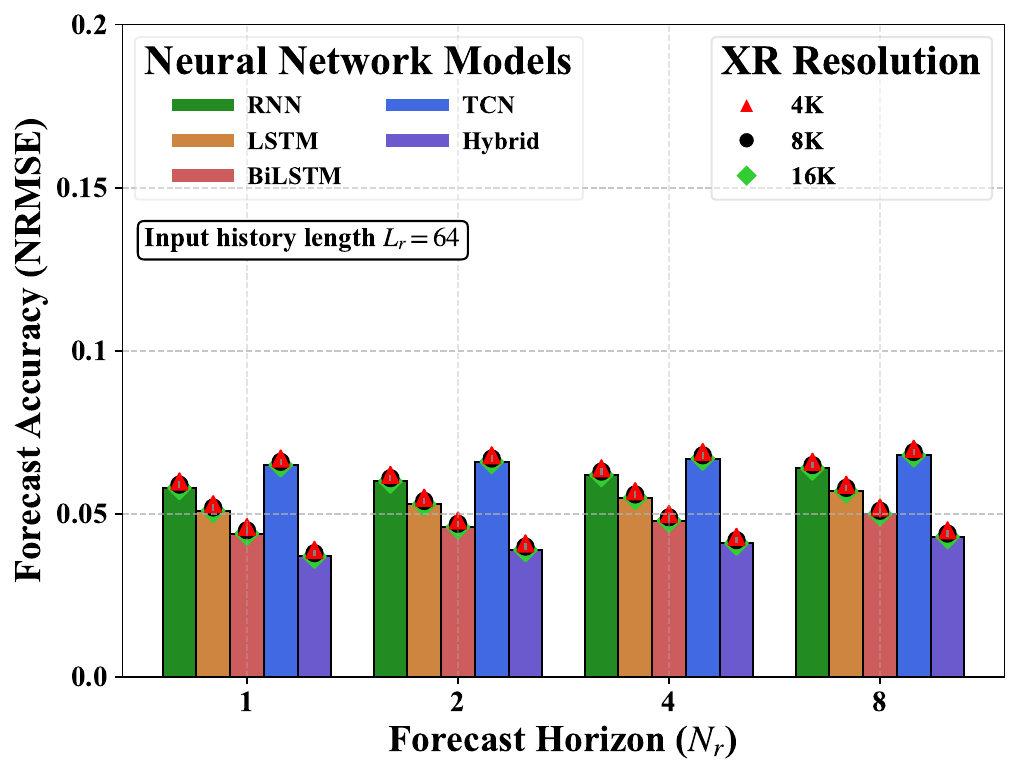}\label{rmsse13}%
  }

  \caption{Comparison of NRMSE of considered time-series forecasts by neural network-based models with training history length of $L_r = 64$ samples for: (a) XR video, (b) HMD orientations, and (c) control and haptic traffic, against multi-sample-ahead forecast horizons.}
  \label{rmsse_pred1}
\end{figure*}
\setlength{\textfloatsep}{1pt}

\begin{figure*}[!t]
  \centering
  \subfloat[XR video]{%
    \includegraphics[width=0.333\textwidth]{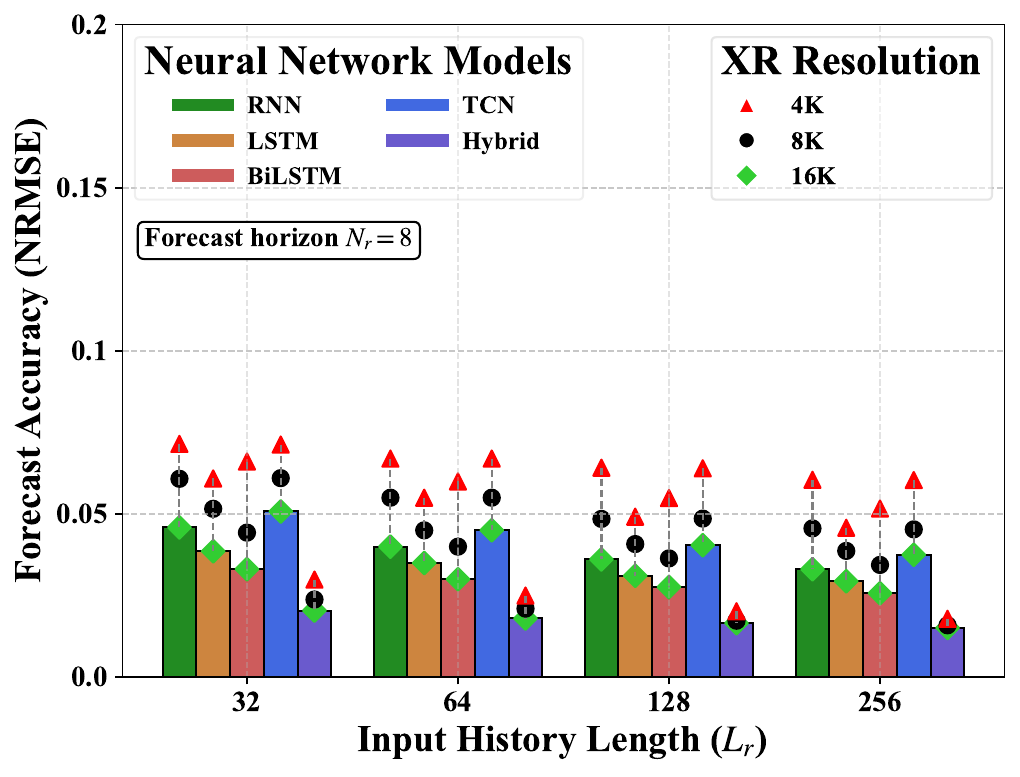}\label{rmsse21}%
  }
  \subfloat[HMD orientations]{%
    \includegraphics[width=0.333\textwidth]{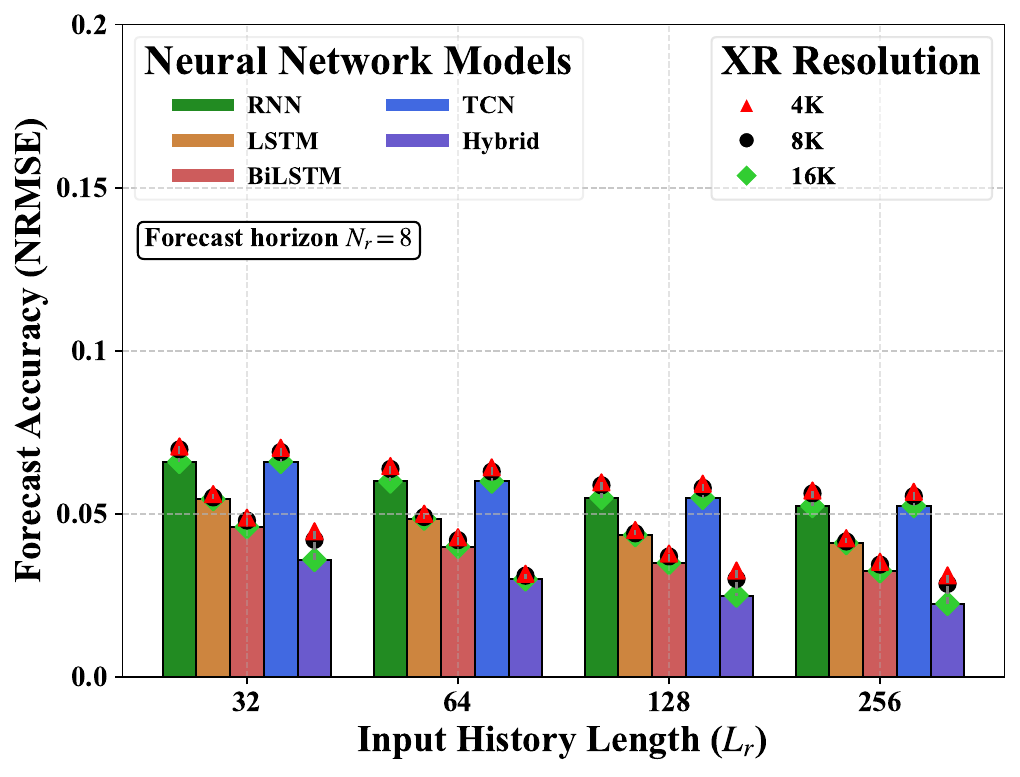}\label{rmsse22}%
  }
  \subfloat[Control and Haptics]{%
    \includegraphics[width=0.333\textwidth]{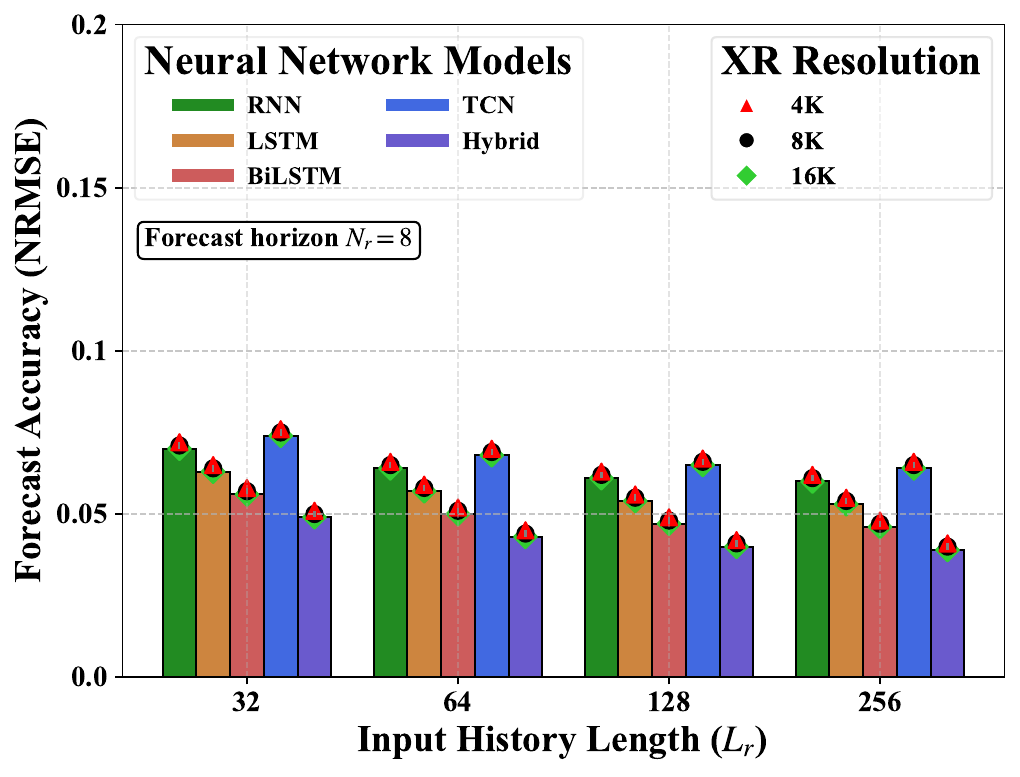}\label{rmsse23}%
  }

  \caption{Comparison of NRMSE of considered time-series forecasts by neural network-based models with forecast length of $N_r = 8$ samples for: (a) XR video, (b) HMD orientations, and (c) control and haptic traffic, against input history lengths.}
  \label{rmsse_pred2}
\end{figure*}
\setlength{\textfloatsep}{1pt}

\begin{figure*}[!t]
  \centering
  \subfloat[XR video]{%
    \includegraphics[width=0.333\textwidth]{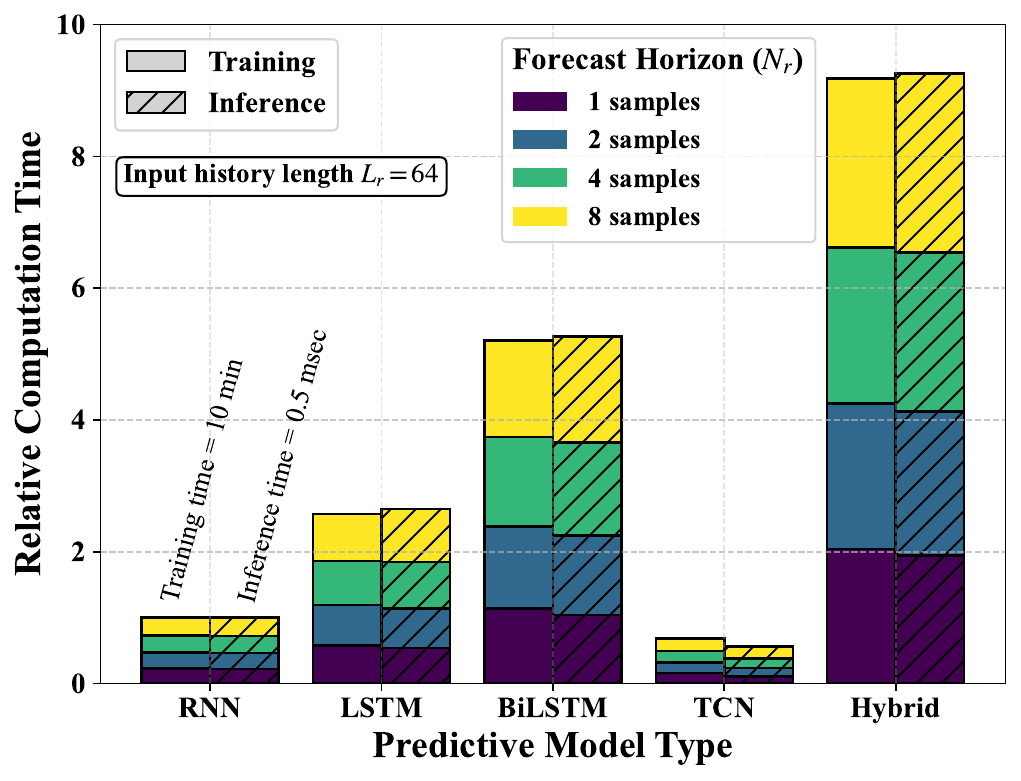}\label{times11}%
  }
  \subfloat[HMD orientations]{%
    \includegraphics[width=0.333\textwidth]{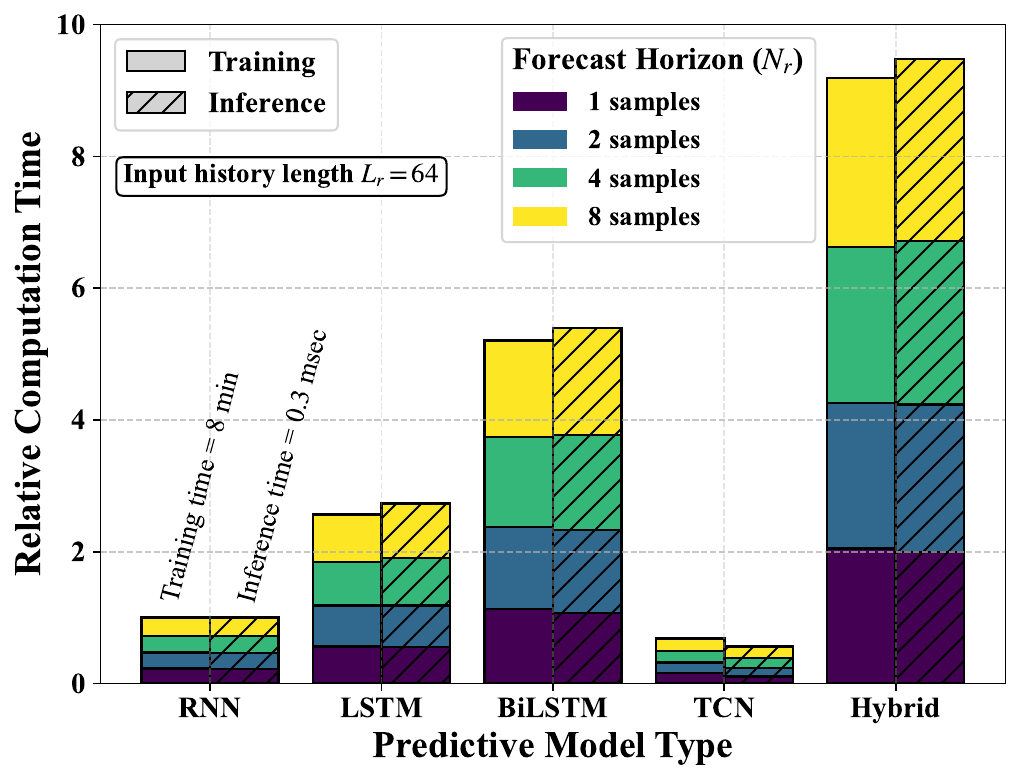}\label{times12}%
  }
  \subfloat[Control and Haptics]{%
    \includegraphics[width=0.333\textwidth]{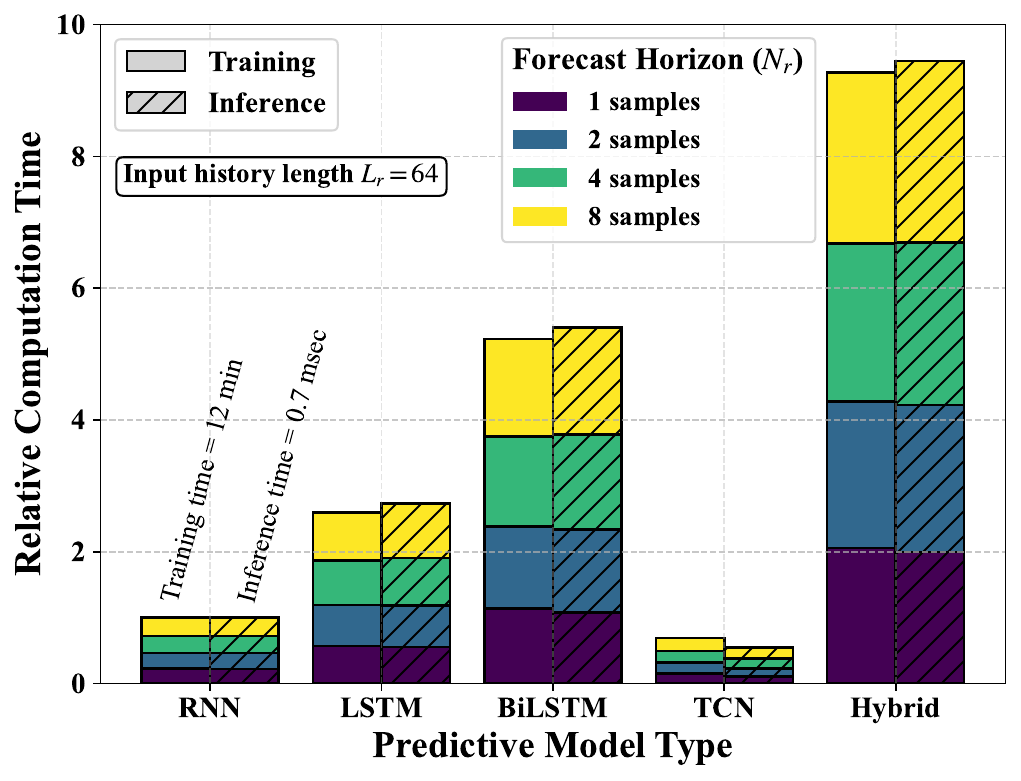}\label{times13}%
  }

  \caption{Comparison of training and inference times of considered neural network-based time-series forecast models for long-horizon multi-sample-ahead forecasts of: (a) XR video, (b) HMD orientations, and (c) control and haptic traffic.}
  \label{times_pred}
\end{figure*}
\setlength{\textfloatsep}{1pt}


\subsection{Results and Discussions} \label{sec6.2}
The OMNeT++ network simulation was executed for 5 seconds per load instance, with the aggregated network load varying from 0.1 to 1.0. This setup facilitated the collection of traffic arrival data at every polling cycle of 125~$\mu$sec from 256 SFUs ($\sim100\times10^6$ samples) and 32 ONUs ($\sim12\times10^6$ samples) with each of 4K, 8K, and 16K XR video qualities. All neural network and Transformer-based traffic forecasting models were implemented using the PyTorch framework, with training and testing conducted on a HP Omen laptop with CPU: Intel Core Ultra 9, GPU: Nvidia RTX 5080, and RAM: 32 GB. Following offline training, the models were integrated into an interactive PyTorch$\leftrightarrow$OMNeT++ co-simulation environment to evaluate the performance of the proactive resource allocation schemes.\par
\begin{figure*}[!t]
  \centering
  \subfloat[input history length $L_h$= 128]{%
    \includegraphics[width=0.333\textwidth]{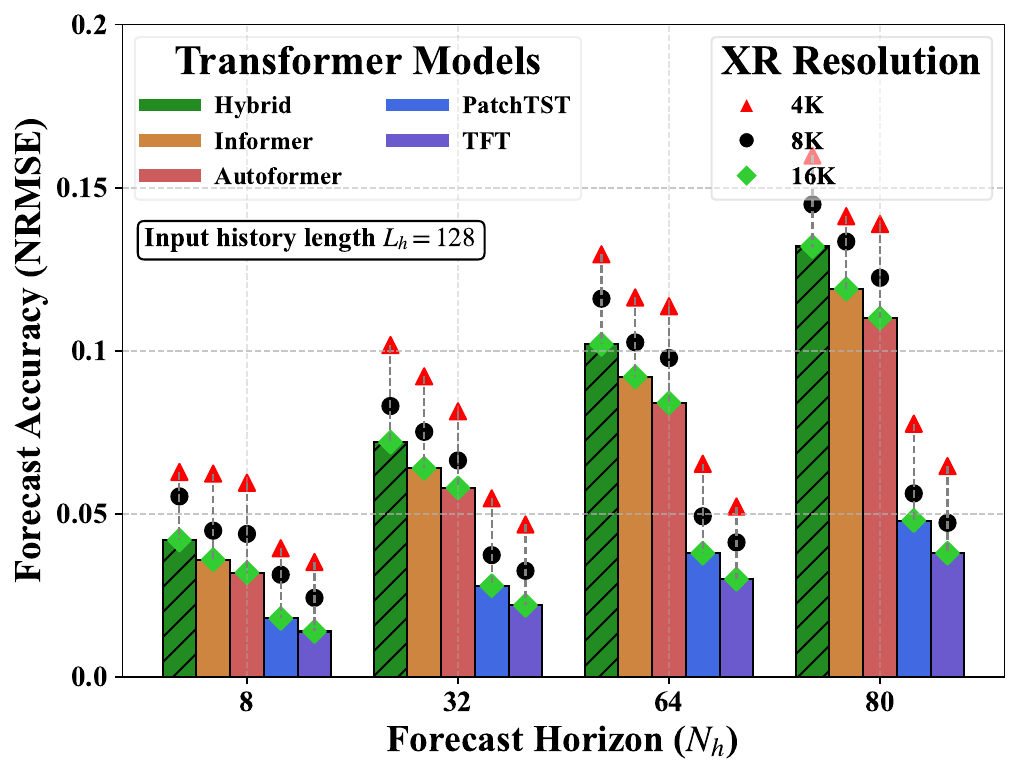}\label{rmsse31}%
  }
  \subfloat[input history length $L_h$ = 256]{%
    \includegraphics[width=0.333\textwidth]{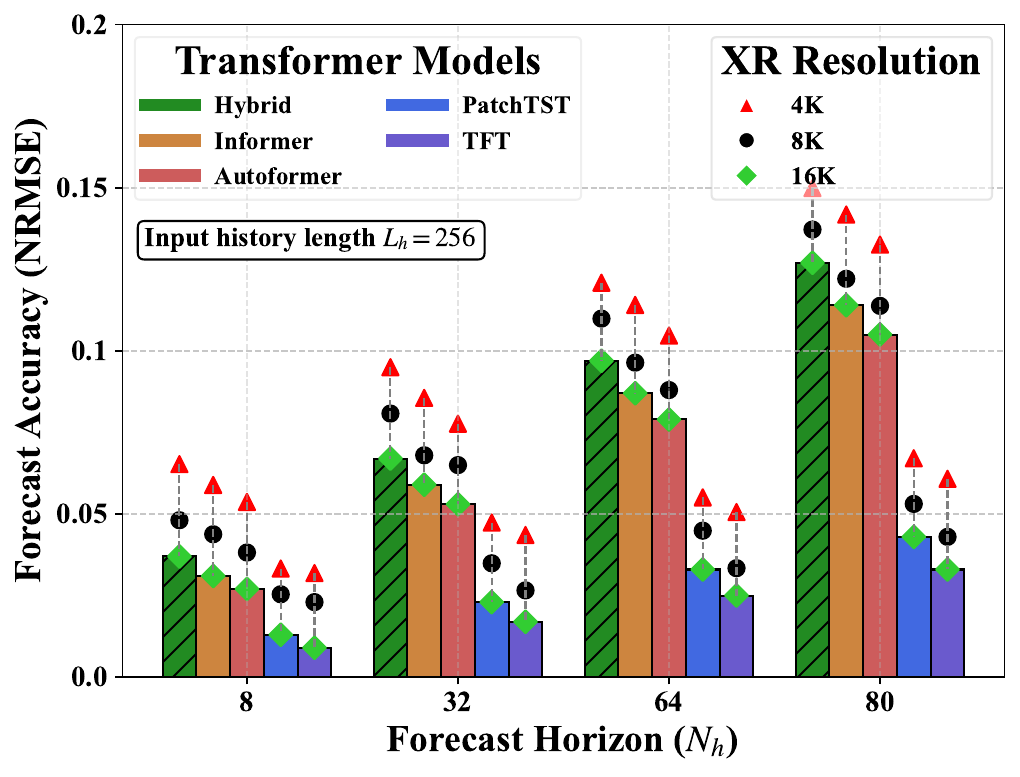}\label{rmsse32}%
  }
  \subfloat[input history length $L_h$ = 512]{%
    \includegraphics[width=0.333\textwidth]{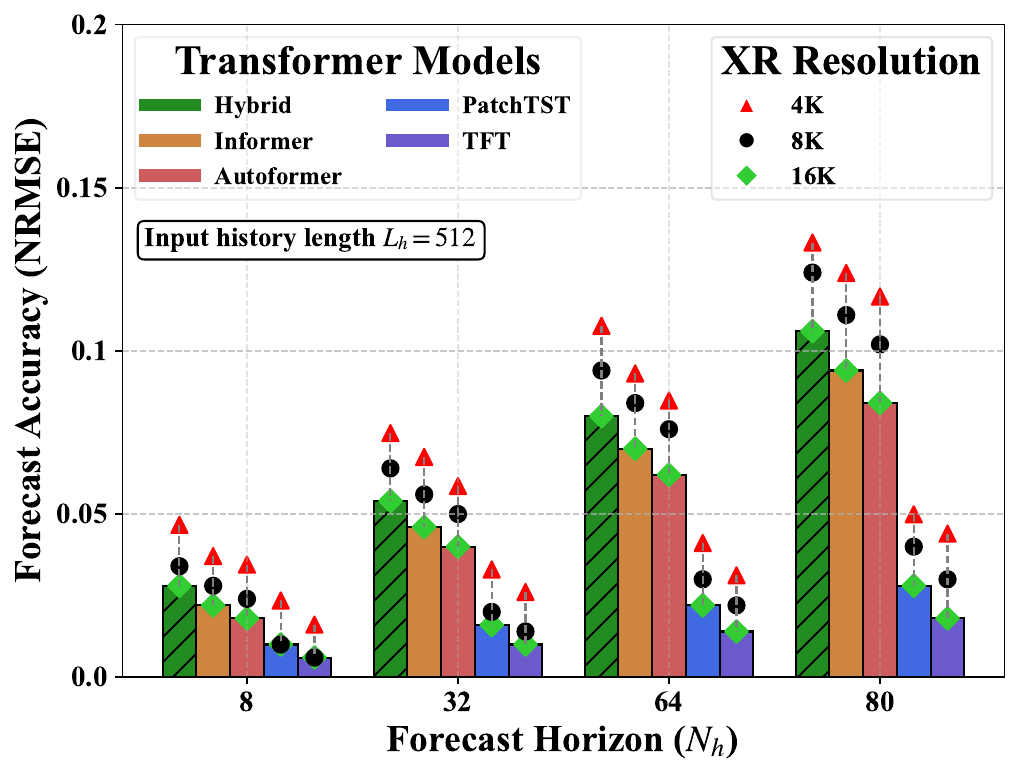}\label{rmsse33}%
  }

  \caption{Comparison of NRMSE of the Transformer-based time-series forecast models and the hybrid TCN-BiLSTM model against long-horizon forecasts of aggregated XR traffic trained with: (a) history length = 128, (b) history length = 256, and (c) history length = 512.}
  \label{rmsse_gen}
\end{figure*}
\setlength{\textfloatsep}{1pt}

\begin{figure*}[!t]
  \centering
  \subfloat[input history length $L_h$ = 128]{%
    \includegraphics[width=0.333\textwidth]{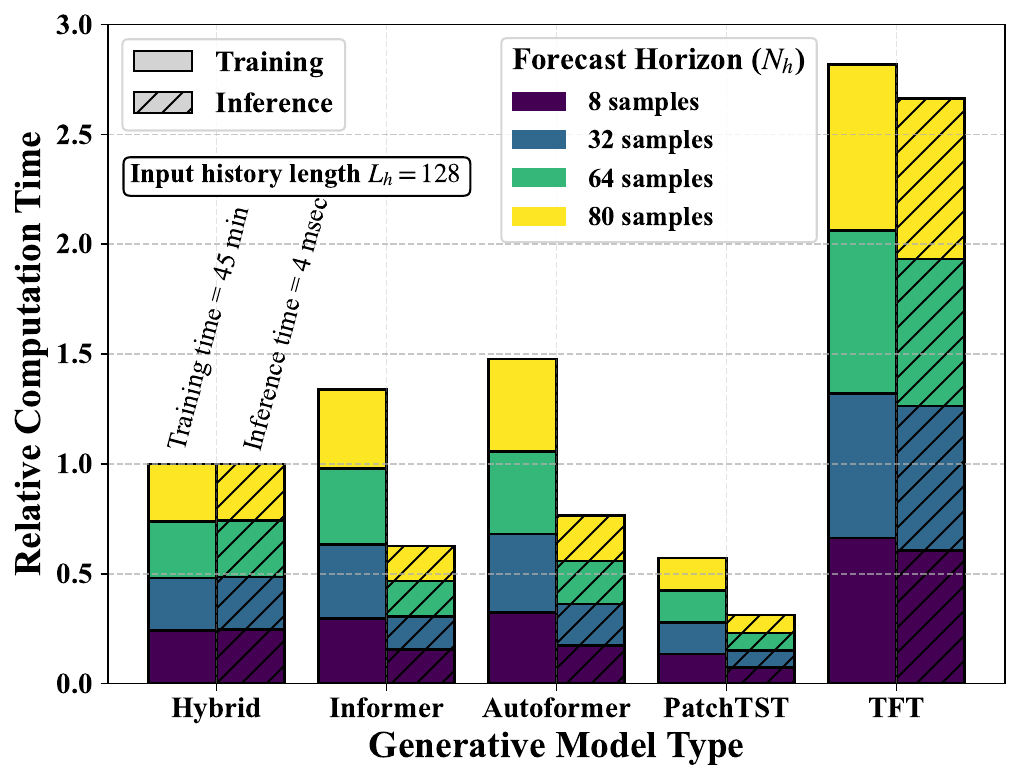}\label{times21}%
  }
  \subfloat[input history length $L_h$ = 256]{%
    \includegraphics[width=0.333\textwidth]{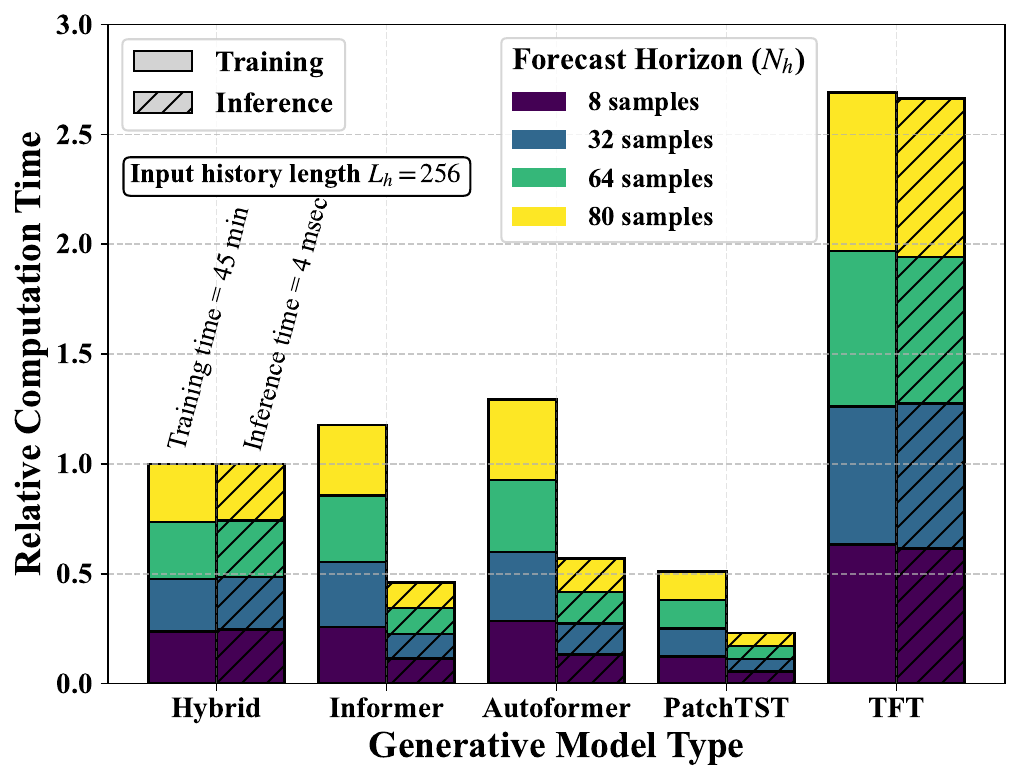}\label{times22}%
  }
  \subfloat[input history length $L_h$ = 512]{%
    \includegraphics[width=0.333\textwidth]{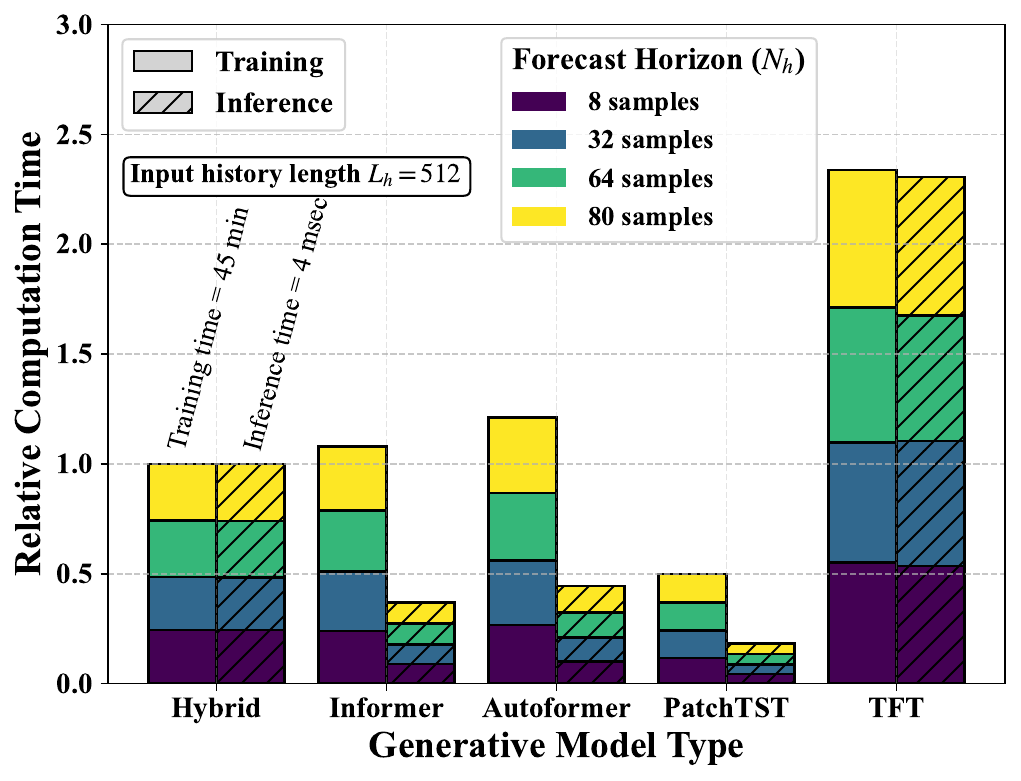}\label{times23}%
  }

  \caption{Comparison of training and inference times of the Transformer-based time-series models and the hybrid TCN-BiLSTM model for long-horizon forecasts of aggregated XR traffic trained with: (a) history length = 128, (b) history length = 256, and (c) history length = 512.}
  \label{times_gen}
\end{figure*}
\setlength{\textfloatsep}{1pt}
%
In Fig.~\ref{rmsse_pred1}, we show the NRMSE variation for the considered predictive models, i.e., RNN, LSTM, BiLSTM, TCN, and hybrid (1TCN+2BiLSTM) across long-horizon multi-sample-ahead forecasting of XR video, HMD orientations, and Control/Haptics traffic. With a fixed history length of $L_r$ = 64, we observe a monotonic increase in NRMSE as the forecast horizon extends from $N_r$ = 1 to 8 (125~$\mu$sec to 1 msec), an expected consequence of error accumulation. While the standard RNN is constrained by the vanishing gradient problem, the proposed hybrid architecture achieves significantly lower error rates than all baselines, including the standalone TCN. This accuracy is attributed to the TCN layer effectively extracting local short-term features, which are subsequently processed by the stacked BiLSTM layers to capture temporal dependencies. Note that higher resolutions (8K and 16K) result in lower relative errors compared to 4K due to denser, more predictable traffic patterns, whereas Control and Haptics traffic demonstrate the highest error due to jitter from Generalized Pareto distributed packet arrival times. These observations highlight the importance of architecture design in accommodating both smooth high-rate traffic and stochastic low-rate control streams.\par
Subsequently, we analyze the impact of input history length and computational complexity on these predictive models. As shown in Fig.~\ref{rmsse_pred2}, increasing the history length $L_r$ generally decreases NRMSE by capturing long-term dependencies and improving robustness against bursts, but extending beyond $L_r$ = 64 yields minimal gains. In terms of complexity, the sequential nature of RNN, LSTM, BiLSTM, and hybrid models results in significantly higher training and inference times compared to the TCN, as shown in Fig.~\ref{times_pred}. Although the hybrid model demands more offline training resources, its superior accuracy justifies the cost. To ensure practical deployment, the starting times of model inference must be adjusted according to Sec.~\ref{sec5.1} to effectively embed them within the FTTR-WiFi proactive allocation schemes.\par
\begin{figure*}[!t]
  \centering
  \subfloat[Objective and computation time comparison]{%
    \includegraphics[width=0.333\textwidth]{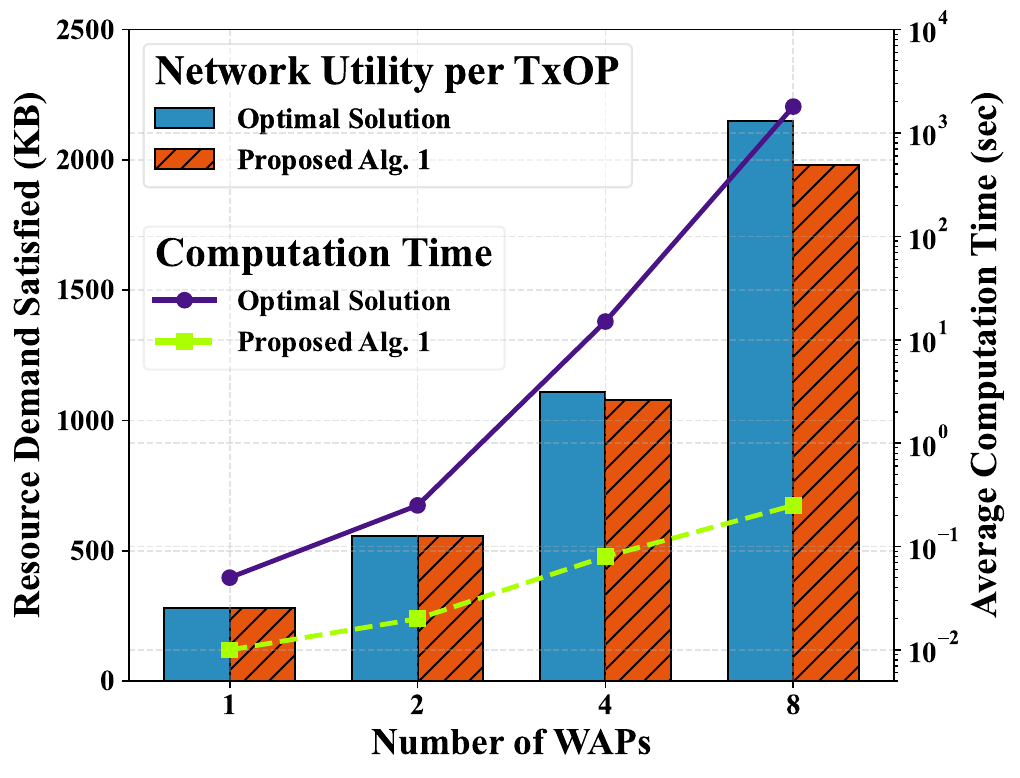}
    \label{wifi1}%
  }
  \subfloat[Wireless interface latency comparison]{%
    \includegraphics[width=0.333\textwidth]{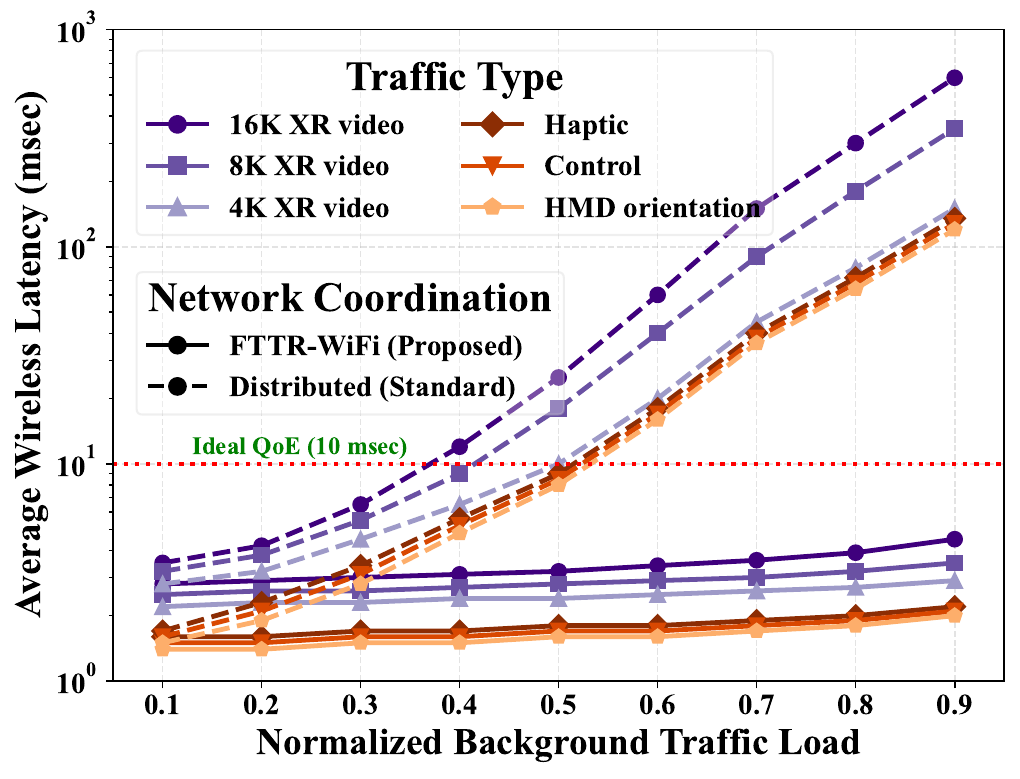}\label{wifi2}%
  }
  \subfloat[Wireless interface latency reduction by handover]{%
    \includegraphics[width=0.333\textwidth]{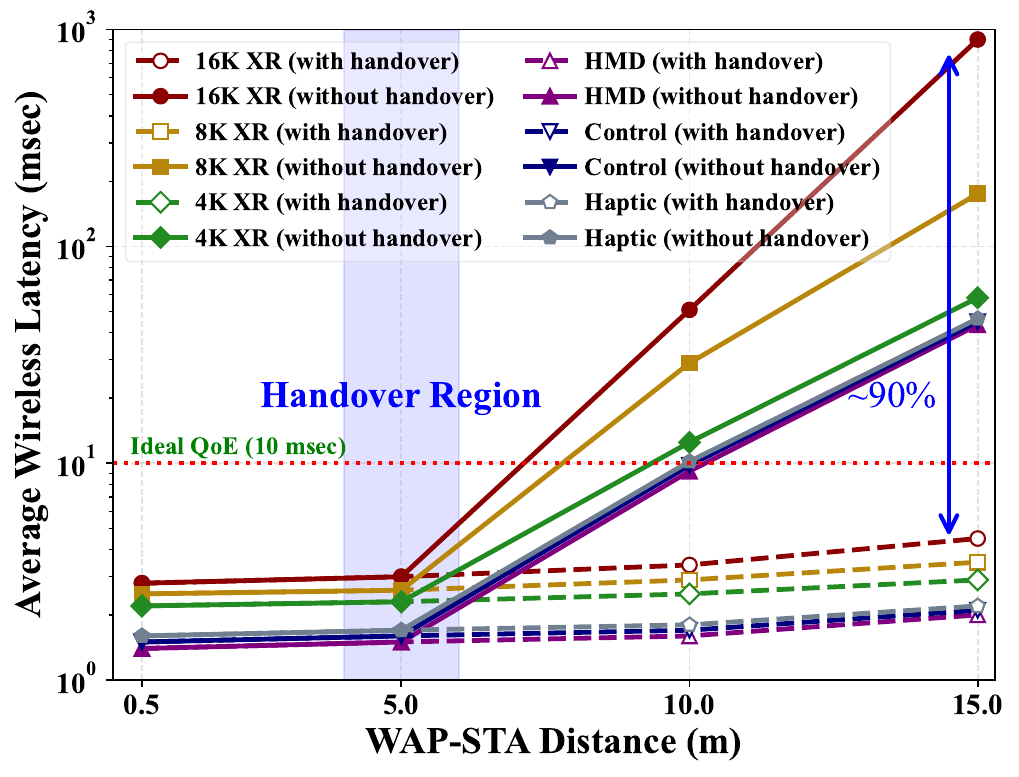}\label{wifi3}%
  }

  \caption{Comparison of: (a) network utilities achieved and computation times consumed by the optimal solution and the proposed Alg.~\ref{alg1}, (b) average wireless interface latency of multi-modal XR traffic with the proposed FTTR-WiFi coordination and standard distributed WiFi, and (c) average wireless interface latency reduction with the proposed network-orchestrated intelligent in-premise handovers.}
  \label{wifi_perf}
\end{figure*}
\setlength{\textfloatsep}{1pt}

\begin{figure*}[!t]
  \centering
  \subfloat[4K XR video for fair QoE]{%
    \includegraphics[width=0.333\textwidth]{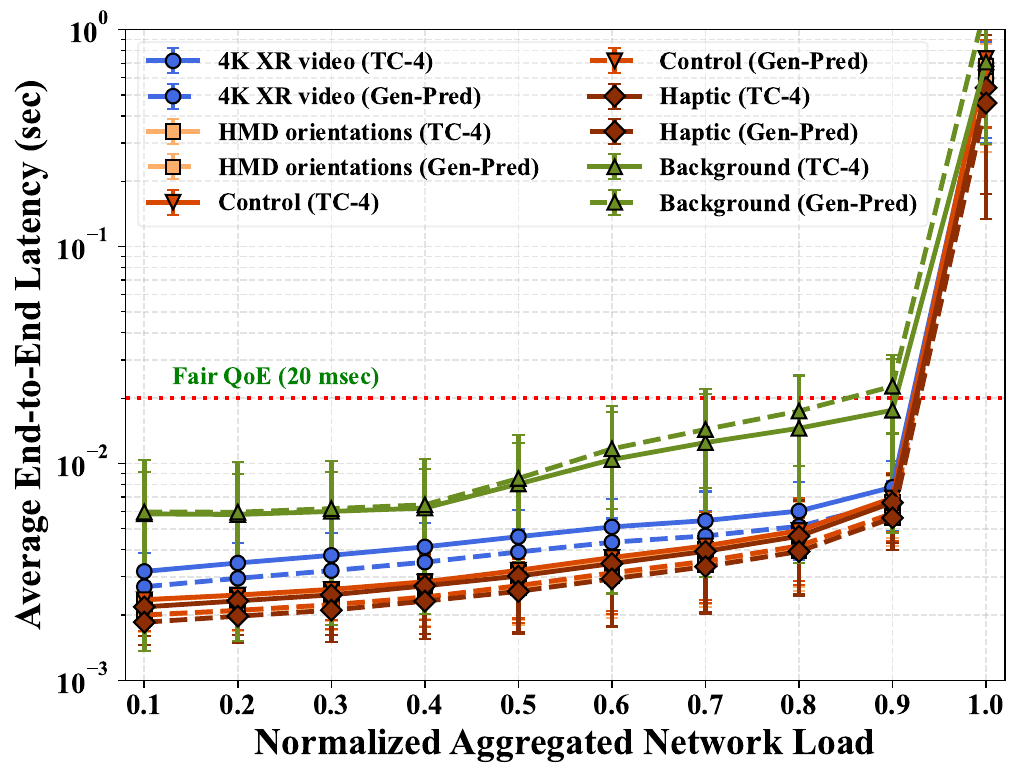}\label{e2e_4k}%
  }
  \subfloat[8K XR video for comfortable QoE]{%
    \includegraphics[width=0.333\textwidth]{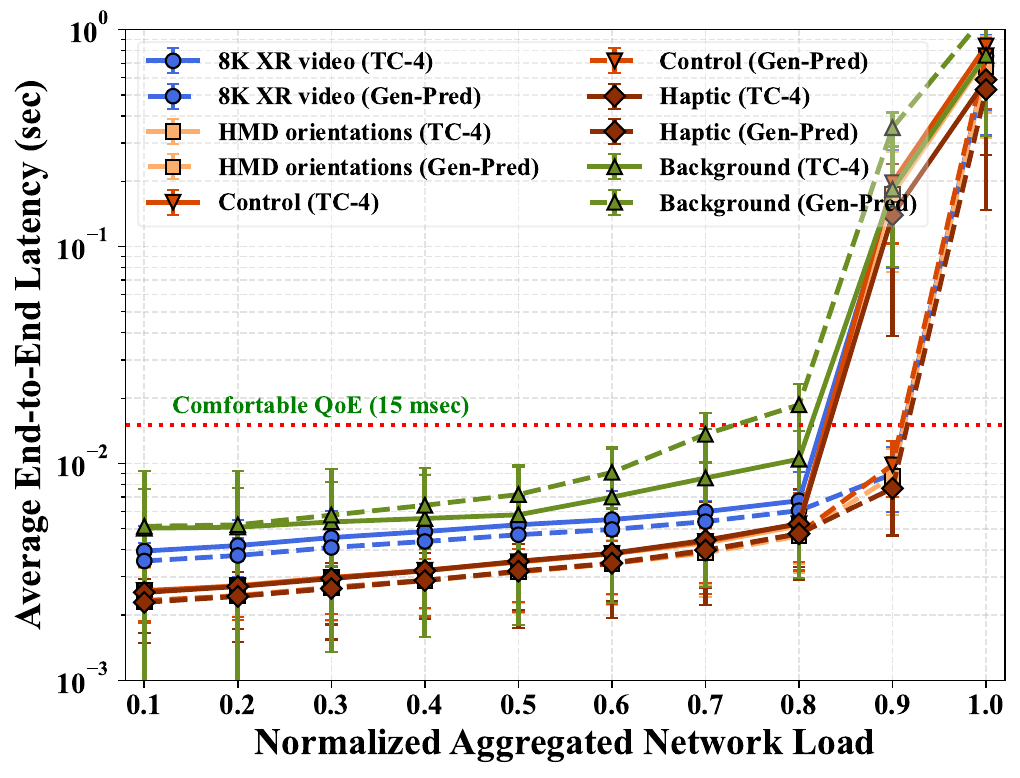}\label{e2e_8k}%
  }
  \subfloat[16K XR video for ideal QoE]{%
    \includegraphics[width=0.333\textwidth]{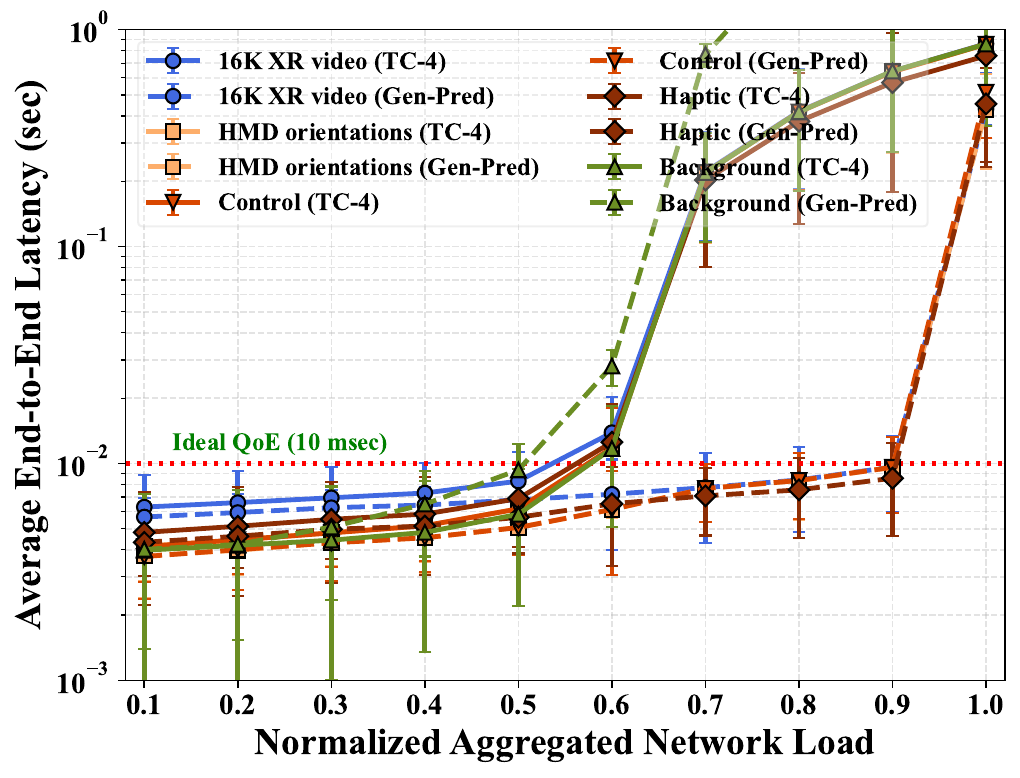}\label{e2e_16k}%
  }

  \caption{Comparison of end-to-end latency of video, HMD orientations, control, haptic, and background traffic in XR collaborations with the proposed resource allocation scheme against G-PON TC-4 baseline scheduling scheme over the integrated FTTH (50G-PON, 1:32), FTTR (XGS-EPON, 1:8), and WiFi-7 while transmitting XR video frames of quality (a) 4K, (b) 8K, and (c) 16K.}
  \label{e2e_latency}
\end{figure*}
\setlength{\textfloatsep}{1pt}
Next, in Fig.~\ref{rmsse_gen}, we evaluate the forecasting accuracy of aggregated multi-modal XR traffic at the ONUs using the considered Transformer-based models, i.e., Informer, Autoformer, PatchTST, and TFT alongside the best predictive hybrid model. As shown in Figs.~\ref{rmsse31}-\ref{rmsse33}, both Informer and Autoformer outperform the hybrid model over long horizons $N_h \in \{8,32,64,80\}$, with PatchTST and TFT achieving further notable ($\sim$75\%) improvements. The hybrid model suffers from sequential error accumulation, whereas Transformer-based models exploit self-attention to capture global temporal structure and benefit from the smoother aggregated waveforms arising from statistical multiplexing. These results confirm that the Transformer-based models provide more stable long-horizon multi-sample-ahead forecasts and substantially lower NRMSE, particularly at longer horizons where the accuracy of sequential models degrade more rapidly.\par
In Fig.~\ref{times_gen}, we also compare the computational cost of the Transformer-based models relative to the hybrid model. Due to global self-attention, most Transformer-based models exhibit higher training times, with PatchTST being the primary exception as its patching mechanism reduces sequence length and overall complexity. In contrast, inference latencies for Informer, Autoformer, and PatchTST are generally lower than the hybrid model because Transformer layers allow parallel evaluation, making them suitable for real-time predictive scheduling. The only outlier is TFT, whose variable-selection and gating networks introduce substantial overhead that results in inference times exceeding those of the hybrid model. This distinction highlights that while Transformers offer strong accuracy-latency trade-offs, not all variants meet stringent real-time constraints, especially when additional interpretability modules are included like in TFT.\par
Efficient STA-WAP associations and seamless in-premise handovers are essential for meeting the end-to-end QoE of multi-user XR. In Fig.~\ref{wifi1}, we demonstrate that Alg.~\ref{alg1} approximates the optimal objective function value or the network utility obtained via Gurobi solver with substantially lower computational time, which is critical for adapting to dynamic mobility. The proposed FTTR-WiFi coordinated framework consistently maintains wireless interface latency within ideal QoE requirements, as shown in Fig.~\ref{wifi2}. However, distributed WiFi networks without coordination is unable to satisfy to these strict QoE requirements. Furthermore, as users move between rooms, the increasing WAP-STA distance degrades channel gain and throughput in turn. Hence, handing over to a neighboring WAP using the proposed strategy significantly mitigates this degradation, as shown in Fig.~\ref{wifi3}, reducing 16K XR frame latency by nearly 90\% compared to maintaining a connection at 15~meter.\par

Finally, in Fig.~\ref{e2e_latency}, we present the end-to-end latency of multi-modal XR traffic over the integrated FTTR architecture. As evidenced in Figs.~\ref{e2e_4k}-\ref{e2e_8k}, the average latencies for 4K and 8K XR video under the default reactive TC-4 resource allocation remain within fair and comfortable QoE requirements, respectively. However, they exhibit significant jitter which degrades the user experience. Conversely, as illustrated in Fig.~\ref{e2e_16k}, the latency for 16K video violates the ideal QoE requirement once the normalized aggregated network load surpasses 0.5. Nevertheless, by adopting our proposed resource allocation framework, the end-to-end latency is minimized and jitter is substantially suppressed across all XR qualities. Crucially, even for 16K XR under heavy load conditions, both average delay and jitter are maintained well within the ideal QoE requirements. These results conclusively establish that efficient proactive resource allocation is indispensable for satisfying the high-throughput, ultra-low-latency, and high-reliability requirements of immersive XR collaborations.\par

\section{Conclusions} \label{sec7}
In this paper, a novel \emph{framework for proactive coarse-grained resource allocation and intelligent roaming across the integrated FTTR network architecture} has been proposed. An NRMSE of less than 0.1 has been achieved for forecasting multi-modal XR traffic using a hybrid neural network comprising 1~TCN and 2~BiLSTM layers. Thorough numerical evaluations have been performed to comparatively establish that this hybrid architecture is the best predictor for the FTTR-WiFi segment over standard RNN, LSTM, and standalone TCN models. Conversely, for the FTTH segment, Transformer-based models, specifically PatchTST and TFT, have been demonstrated to outperform the hybrid model in long-horizon multi-step-ahead forecasting of complex aggregated traffic. Moreover, utilizing the proposed network-orchestrated seamless roaming strategy, the XR frame transmission latency is reduced by nearly 90\% compared to maintaining connections at large WAP-STA distances. This ensures the wireless interface latency remains within the ideal QoE requirements. By incorporating these accurate traffic forecasts into the proactive resource allocation framework, results show that high-throughput, low-latency, and high-reliability requirements for XR collaborations can be satisfied, even under high network load conditions which otherwise cannot be achieved with current resource allocation schemes.

\appendices
\section{Neural Network-based Forecasting in FTTR Segment} \label{sec.A}
Recent research has established that multi-modal XR traffic has short-term dynamics as well as long-term non-linear temporal dependencies \cite{xr_iotj}. As statistical time-series forecast models, e.g., auto-regressive integrated moving average (ARIMA) and its variants, primarily rely on linear temporal dependencies, their accuracy significantly degrades during long-horizon forecasts. Thus, we adopt deep learning-based time-series forecast models, progressing from fundamental recurrent structures to more evolved convolutional-recurrent hybrids. \emph{Each forecast model consists of a single temporal encoding layer using TCN, RNN, LSTM, or BiLSTM and a dense output layer that produces $(N_f+N_r)$ multi-step forecasts}. Finally, we extend this to \emph{a hybrid architecture with 1~TCN and 2 stacked BiLSTM layers}. The simplest model for processing sequential data is the \textbf{RNN model}, which maintains a ``memory" or ``state" of past information, updated at each time step \cite{rnn_book}. For an input element $x_k$ from the sequence $\mathbf{x}_{t-N_f}$ (omitting superscript $(i,j)$ for readability), the hidden state $\bm{h}_k, \forall k \in [t-N_f-L_r+1, t-N_f]$ and final forecast vector $\hat{\bm{y}}_t\in\mathbb{R}^{N_f+N_r}$ are calculated as:\\
\vspace{-\baselineskip}
\begin{align}
\bm{h}_k &= \varphi(\bm{W}_{hh} \bm{h}_{k-1} + \bm{W}_{xh} x_k + \bm{b}_h), \label{eqRNN1}\\
 \hat{\bm{y}}_{t-N_f} &= \bm{W}_{hy} \bm{h}_{t-N_f} + \bm{b}_y, \label{eqRNN2}
\end{align}
where $\bm{W}_{hh}$, $\bm{W}_{xh}$, and $\bm{W}_{hy}$ are weight matrices, $\bm{b}_h$ and $\bm{b}_y$ are bias terms, $\varphi(\cdot)$ is $\tanh(.)$, and $\bm{h}_{t-N_f}$ is the final hidden state after processing the entire $L_r$-length input sequence. Although RNNs are simple to implement, the gradient decay or explode problem while training with backpropagation through time (BPTT) over long sequences limits its ability to capture long-range temporal dependencies.\par
To alleviate these limitations, the \textbf{LSTM model} introduces \emph{memory cell} and \emph{gating mechanisms}. The cell state acts as an unimpeded memory pathway, while gates regulate information flow. The primary equations of an LSTM cell at time step $k$ (for input $x_k$) and the final forecast vector are \cite{lstm_comp}:\\
\vspace{-\baselineskip}
\begin{align}
 \bm{i}_k &= \sigma(\bm{W}_{ih} \bm{h}_{k-1} + \bm{W}_{ix} x_k + \bm{b}_i), \forall k, \label{lstm1}\\
 \bm{f}_k &= \sigma(\bm{W}_{fh} \bm{h}_{k-1} + \bm{W}_{fx} x_k + \bm{b}_f), \forall k, \label{lstm2}\\
 \bm{o}_k &= \sigma(\bm{W}_{oh} \bm{h}_{k-1} + \bm{W}_{ox} x_k + \bm{b}_o), \forall k, \label{lstm3}\\
 \bm{\tilde{c}}_k &= \tanh(\bm{W}_{ch} \bm{h}_{k-1}+ \bm{W}_{cx} x_k + \bm{b}_c), \forall k, \label{lstm4}\\
 \bm{c}_k &= \bm{f}_k \odot \bm{c}_{k-1} + \bm{i}_k \odot \tilde{\bm{c}}_k, \forall k, \label{lstm5}\\
 \bm{h}_k &= \bm{o}_k \odot \tanh(\bm{c}_k), \forall k, \label{lstm6}\\
 \hat{\bm{y}}_{t-N_f} &= \bm{W}_{hy} \bm{h}_{t-N_f} + \bm{b}_{y}. \label{lstm7}
\end{align}
where $\bm{i}_k$, $\bm{f}_k$, $\bm{o}_k$ are the gate parameters, $\bm{W}_{ih}$, $\bm{W}_{ix}$, $\bm{W}_{fh}$, $\bm{W}_{fx}$, $\bm{W}_{oh}$, $\bm{W}_{ox}$, $\bm{W}_{hy}$ are the weight matrices, $\bm{b}_h$ and $\bm{b}_y$ are bias terms, $\bm{c}_k$ is the cell state, $\bm{h}_k$ is the cell output, $\varphi(\cdot)$ is a non-linear activation function, typically $\tanh(.)$, and $\hat{\bm{y}}_{t-N_f}$ is generated from the final hidden state $\bm{h}_{t-N_f}$. Nevertheless, as the LSTM nodes process information only in the forward direction, its performance may be limited. A \textbf{BiLSTM model} addresses this by processing the input sequence $\mathbf{x}_{t-N_f}$ from both directions in parallel. It consists of two independent LSTM layers: the \textit{forward LSTM} processes the sequence from $x_{t-N_f-L_r+1}$ to $x_{t-N_f}$, generating a sequence of forward hidden states $[\overrightarrow{\bm{h}}_{t-N_f-L_r+1}, \dots, \overrightarrow{\bm{h}}_{t-N_f}]$, and the \textit{backward LSTM} processes the sequence in reverse, from $x_{t-N_f}$ down to $x_{t-N_f-L_r+1}$, generating a sequence of backward hidden states $[\overleftarrow{\bm{h}}_{t-N_f-L_r+1}, \dots, \overleftarrow{\bm{h}}_{t-N_f}]$. At any time step $k$, the hidden state $\bm{h}_k$ is the concatenation of the forward and backward states, i.e., $\bm{h}_k = [\overrightarrow{\bm{h}}_k;\overleftarrow{\bm{h}}_k]$. The final hidden state $\bm{h}_{t-N_f}$ at the end of the input history is used by the output dense layer to generate the $(N_f+N_r)$-step forecast vector $\hat{\bm{y}}_{t-N_f} = \bm{W}_{hy} \bm{h}_{t-N_f} + \bm{b}_{y}$.\par
Unlike recurrent networks, a \textbf{TCN model} adopts a non-recurrent architecture that applies a set of $F$ learnable filters or kernels $\bm{w} \in \mathbb{R}^M$ across the input sequence $\mathbf{x}_{t-N_f}$ \cite{tcn}. For each single filter $\bm{w}$, a feature map $\bm{c}$ is produced, whose $m$-th element $c_m$ captures a specific local temporal pattern and is calculated as $c_m = \varphi\left(b + \sum_{n=0}^{M-1} w_{n+1} \cdot x_{t-N_f-L_r+m+n}\right)$ where $\varphi(\cdot)$ is a non-linear activation function (typically ReLU) and $b$ is a bias term. The combined output of all filters forms the feature tensor $\bm{H}_{conv}$. A global max-pooling layer then captures the maximum activation value across the full length of each feature map in $\bm{H}_{conv}$, condensing the most salient temporal features into a compact feature vector $\bm{h}_{pool} \in \mathbb{R}^F$. This vector is subsequently passed to the final dense output layer. The key equations for the TCN are:\\
\vspace{-\baselineskip}
\begin{align}
 \bm{H}_{conv} &= \varphi(\text{Conv1D}(\mathbf{x}_{t-N_f})), \label{cnn1}\\
 \bm{h}_{pool} &= \text{GlobalMaxPool}(\bm{H}_{conv}), \label{cnn2} \\
 \hat{\bm{y}}_{t-N_f} &= \bm{W}_{hy} \bm{h}_{pool} + \bm{b}_{y}. \label{cnn3}
\end{align}
\par Although the TCN is computationally efficient, its ability to model long-term temporal dependencies is generally weaker than that of LSTM or BiLSTM networks. To leverage the complementary strengths of both architectures, we construct a \textbf{hybrid model} by stacking one TCN layer followed by two BiLSTM layers. The TCN layer acts as a feature extractor, passing its output sequence $\bm{H}_{conv}$ without pooling to the BiLSTM layers, as expressed as follows:\\
\vspace{-\baselineskip}
\begin{align}
 \bm{H}_{conv} &= \varphi(\text{Conv1D}(\mathbf{x}_{t-N_f})), \\
 \bm{H}_{bilstm}^{(1)} &= \text{BiLSTM}_1(\bm{H}_{conv}), \\
 \bm{H}_{bilstm}^{(2)} &= \text{BiLSTM}_2(\bm{H}_{bilstm}^{(1)}), \\
 \bm{h}_{final}^{(2)} &= \text{LastState}(\bm{H}_{bilstm}^{(2)}), \\
 \hat{\bm{y}}_{t-N_f} &= \bm{W}_{hy} \bm{h}_{final}^{(2)} + \bm{b}_{y},
 \label{Hybrid}
\end{align}
where $\bm{H}_{conv} \in \mathbb{R}^{L_r' \times F}$ is the feature sequence from the TCN layer, $\bm{H}_{bilstm}^{(1)}$ and $\bm{H}_{bilstm}^{(2)}$ are the output sequences from the BiLSTM layers, and $\bm{h}_{final}^{(2)}$ is the final concatenated hidden state corresponding to timeslot-$(t-N_f)$.\par
It is critical to note that the performance of all the aforementioned models relies on a careful selection of hyperparameters. Each model is trained using the \emph{Adam optimizer} to minimize the MSE loss. Essential hyperparameters optimized include: the learning rate, e.g., $10^{-8}$ to $10^{-2}$, the batch size $B \in \{32, 64, 128\}$, and the dropout probability $p_d \in \{0.1, 0.2, 0.5\}$ for regularization. The number of hidden units in the recurrent layer(s) of RNN, LSTM, and BiLSTM are selected from $\{32, 64, 128, 256\}$. The number of filters $F$ per layer is the same as the number of recurrent units and the kernel size $M \in \{3, 5, 7\}$. The input history length $L_r$ is set to be significantly larger than $(N_f+N_r)$, e.g., $L_r \in \{32,64,128, 256, 512\}$. The \emph{computational complexity} per training iteration primarily depends on the hidden layer size $H$, input length $L_r$, and forecast horizon $(N_f+N_r)$ \cite{rnn_book}. The RNN and LSTM require $\mathcal{O}(L_r H^2)$ operations, while the BiLSTM doubles this cost. The TCN scales as $\mathcal{O}(L_r M F)$ and benefits from high parallelism. The hybrid 1~TCN+2~BiLSTM model has the highest complexity $\mathcal{O}(L_r(MF + 2H^2))$, while also achieving the best performance in traffic forecasting.

\section{Transformer-based Forecasting in FTTH Segment} \label{sec.B}
Transformers were originally designed for natural language processing but their impressive performance in sequence learning problems motivated their adoption in diverse domains, especially those requiring long-horizon time-series forecasting. The \emph{canonical self-attention vector} is calculated as \cite{bishop2023deep}:\\
\vspace{-\baselineskip}
\begin{align}
\mathcal{A}_{\text{Self}}(\bm{Q},\bm{Z},\bm{V}) = \text{Softmax}\left(\frac{\bm{Q} \bm{Z}^\top}{\sqrt{d_z}}\right)\bm{V},
\end{align}
where $\bm{Q}\in\mathbb{R}^{n\times d_z}$, $\bm{Z}\in\mathbb{R}^{m\times d_z}$, $\bm{V}\in\mathbb{R}^{m\times d_v}$ represent the matrix representations for $n$ queries and $m$ key-value pairs, respectively; $d_z$ denotes the dimension of queries (and keys), and $d_v$ denotes the dimension of values. Note that, in the current context of the encoder, the sequence length is the history length $L_h$, thus $\bm{Q}$, $\bm{Z}$, and $\bm{V}$ are all derived from the historical input data $\bm{\xi}_{t-N_g}$, making $n=L_h$ and $m=L_h$. The matrices are formed by $\bm{Q} = \bm{\xi}\bm{W}^Q$, $\bm{Z} = \bm{\xi}\bm{W}^Z$, $\bm{V} = \bm{\xi}\bm{W}^V$. However, standard Transformers suffer from critical limitations: (a) \emph{quadratic complexity}: the self-attention mechanism computes $\mathcal{O}(L_h^2)$ dot products, making it computationally infeasible for very long histories $L_h \gg N_h$, (b) \emph{high memory usage}: storing the $(L_h \times L_h)$ attention matrix is prohibitive, and (c) \emph{architectural rigidity}: the encoder-decoder structure forces all temporal information to be compressed into a fixed-length latent representation before decoding. To address these challenges, several efficient Transformer variants are proposed.

The \textbf{Informer model} introduces two key components to tackle the complexity problem \cite{informer}. First, it proposes \emph{ProbSparse self-attention} that reduces the complexity from $\mathcal{O}(L_h^2)$ to $\mathcal{O}(L_h \log L_h)$. Instead of evaluating attention scores for every query-key pair, only the $N_{active} = c \cdot \log L_h$ dominant queries exerting the greatest influence on the attention distribution are selected. The dominance is determined by a \emph{sparsity measure} given by:\\
\vspace{-\baselineskip}
\begin{align}
\mathcal{M}(\bm{q}_m, \mathbf{Z}) = \max_n \left\{ \frac{\bm{q}_m \bm{z}_n^\top}{\sqrt{d_z}} \right\} - \frac{1}{L_h} \sum_{n=1}^{L_h} \frac{\bm{q}_m \bm{z}_n^\top}{\sqrt{d_z}}, \label{eqTr1}
\end{align}
which captures how much a query deviates from the average similarity pattern. Queries with large $\mathcal{M}$ values are retained to form a reduced query matrix $\bar{\bm{Q}}$ and the resulting sparse attention operation is calculated as:\\
\vspace{-\baselineskip}
\begin{align}
\mathcal{A}_{\text{ProbSparse}}(\bm{Q}, \bm{Z}, \bm{V}) = \text{Softmax}\left(\frac{\bar{\bm{Q}} \bm{Z}^\top}{\sqrt{d_z}}\right)\mathbf{V}.
\end{align}
Second, an \emph{attention distillation} layer is employed to progressively shorten the sequence length in the encoder. This is achieved by applying a 1D convolutional layer followed by max-pooling, e.g., $\mathbf{Z}^{\lambda}_{distill} = \text{MaxPool}(\text{ELU}(\text{Conv1D}(\mathbf{Z}^{\lambda-1})))$, effectively halving the sequence length $L_h$ at each layer-$\lambda$ and reducing the memory footprint. Finally, the generative decoder directly outputs the entire $(N_g+N_h)$-step forecast in one shot, avoiding the error accumulation inherent in auto-regressive decoding.\par
The \textbf{Autoformer model} introduces a decomposition architecture combined with a novel auto-correlation mechanism for long-term time series forecasting \cite{autoformer}. It incorporates time-series decomposition as an intrinsic block, using a fixed-window moving average ($\text{AvgPool}$ with padding) within each layer to progressively extract and refine components. For a given input sequence $\bm{\xi}^{(u)}_{t-N_g} \in \mathbb{R}^{L_h}$, the trend-cyclical ($\bm{\xi}_{tr}$) and seasonal ($\bm{\xi}_{sn}$) parts are defined as:\\
\vspace{-\baselineskip}
\begin{align}
    \bm{\xi}_{tr} = \text{AvgPool}(\text{Padding}(\bm{\xi})) \quad \text{and} \quad \bm{\xi}_{sn} = \bm{\xi} - \bm{\xi}_{tr}.
\end{align}
The core innovation is the auto-correlation mechanism, which replaces dot-product-based self-attention. This mechanism efficiently measures period-based dependencies by computing the cross-correlation $\mathcal{R}_{\bm{Q}\bm{Z}}(\tau)$ between queries $\bm{Q}$ and keys $\bm{Z}$ in the frequency domain using the Fast Fourier Transform ($\mathcal{F}$):\\
\vspace{-\baselineskip}
\begin{align}
    \mathcal{R}_{\bm{Q}\bm{Z}}(\tau) = \mathcal{F}^{-1}((\mathcal{F}(\bm{Q}) \otimes \mathcal{F}^*(\bm{Z}))),
\end{align}
where $\otimes$ is element-wise multiplication and $\mathcal{F}^*$ is the conjugate $\mathcal{F}$. The model identifies the top-$L_h'$ most correlated lags, $\mathcal{P} = \{\tau_1, \tau_2, \ldots, \tau_{L_h'}\}$, where $\tau_i = \arg\max_{\tau} \mathcal{R}_{\bm{Q}\bm{Z}}(\tau)$. The final \emph{auto-correlation Attention} aggregates the corresponding values $\bm{V}$ using the attention output as:\\
\vspace{-\baselineskip}
\begin{align}
    \mathcal{A}_{\text{auto-corr}}(\bm{Q},\bm{Z},\bm{V}) = \sum_{i=1}^{L_h'} \bm{M}_{i} \cdot \text{Roll}(\bm{V}, \tau_i),
\end{align}
where $\text{Roll}(\bm{V}, \tau_i)$ shifts the series $\bm{V}$ by the detected lag, and $\bm{M}_i$ is the attention weight matrix derived from the normalized correlation value $\mathcal{R}_{\bm{Q}\bm{Z}}(\tau_i)$. This mechanism achieves an $\mathcal{O}(L_h \log L_h)$ complexity.\par
The \textbf{PatchTST model} borrows its core concept from computer vision \cite{patchTST}, reformulating the forecast as a patch-level representation problem. As we consider separate models for each ONU-$u$, the univariate input $\bm{\xi}^{(u)}_{t-N_g} \in\mathbb{R}^{L_h}$ is segmented into $N_p = L_h/P$ patches, where $P$ is the patch length. The $m$-th patch $\bm{P}_m$ (for $m \in \{1, \dots, N_p\}$) is defined as:\\
\vspace{-\baselineskip}
\begin{align}
\bm{P}_m = [\xi^{(u)}_{t-N_g-L_h + (m-1)P + 1}, \ldots, \xi^{(u)}_{t-N_g-L_h + mP}].
\end{align}
Each patch is then linearly projected into an embedding $\bm{e}_m$ using a shared weight matrix $\bm{W}_P$ to create a token:\\
\vspace{-\baselineskip}
\begin{align}
\bm{e}_m = \bm{W}_P \cdot \text{vec}(\bm{P}_m),
\end{align}
where $\text{vec}(\cdot)$ flattens the patch data. These embeddings form the input token sequence $\bm{E} = [\bm{e}_1, \dots, \bm{e}_{N_p}]$ which is processed by a standard Transformer encoder. The self-attention mechanism thus computes relationships between time patches rather than between time steps. The primary benefit of this patch-level attention is computational efficiency. By operating over $N_p$ patches instead of $L_h$ timesteps, the complexity is substantially reduced to $\mathcal{O}((L_h/P)^2)$, enabling the model to process significantly longer historical sequences. The final output of the encoder is mapped by a linear head to directly generate the multi-step forecast $\hat{\bm{\xi}}^{(u)}_{t-N_g+1:t+N_h}$.\par
Interestingly, the \textbf{TFT model} is adopted to combine the benefits of local processing with long-term dependency modeling \cite{TFT}. Unlike standard Transformers, TFT utilizes a sequence-to-sequence LSTM encoder-decoder structure to process the historical traffic $\bm{\xi}^{(u)}_{t-N_g}$, preserving local temporal context. To capture global dependencies across the look-back window $L_h$, it employs an \emph{interpretable temporal attention} mechanism. This layer specifically relates the future-step queries $\bm{Q}$ of the decoder (generated for the forecast horizon $N_g+N_h$) with the historical keys $\bm{Z}$ and values $\bm{V}$ of the encoder. The interpretable temporal attention vector is computed as:\\
\vspace{-\baselineskip}
\begin{align}
\mathcal{A}_{\text{Interp}}(\bm{Q},\bm{Z},\bm{V}) = \text{Softmax}\left(\frac{\bm{Q} \bm{Z}^\top}{\sqrt{d_z}}\right)\bm{V},
\end{align}
where $\bm{Q} \in \mathbb{R}^{(N_g+N_h) \times d_z}$ represents the decoder states for the prediction horizon, and $\bm{Z}, \bm{V} \in \mathbb{R}^{L_h \times d_z}$ represent the encoder's historical states. By sharing value weights across heads, TFT provides interpretability regarding which time steps in $L_h$ contribute most to the forecast. This results in a complexity of $\mathcal{O}((N_g+N_h) \times L_h)$. The attention output is gated via a gated linear unit (GLU) and processed by a gated residual network (GRN), finally yielding the multi-step traffic forecast $\hat{\bm{\xi}}^{(u)}_{t-N_g+1:t+N_h}$.

\bibliographystyle{IEEEtran}
\bibliography{IEEEabrv,references}
\begin{IEEEbiographynophoto}{Sourav Mondal}
(GS'16–M'21) received PhD from the University of Melbourne in 2020, M.Tech from Indian Institute of Technology Kharagpur in 2014, and B.Tech from Kalyani Govt. Engineering College, affiliated to West Bengal University of Technology in 2012. Currently, he is working as a Research Fellow in the Faculty of Engineering and IT, Department of Electrical and Electronic Engineering of the University of Melbourne. Prior to this, he briefly worked as an EDGE/Marie Skłodowska-Curie Post-doctoral Fellow in Trinity College Dublin, Ireland and as a Senior Lead Systems Engineer (Corporate R\&D) in Qualcomm Bengaluru Design Center, India.
\end{IEEEbiographynophoto} 


\begin{IEEEbiographynophoto}{Elaine Wong}
(M'02–SM'14) received B.E. and Ph.D. degrees from the University of Melbourne, Australia. She is currently the Pro Vice-Chancellor (People \& Equity) and Redmond Barry Distinguished Professor at the Faculty of Engineering and Information Technology, The University of Melbourne. Her research interests include energy-efficient optical and wireless networks, optical-wireless integration, broadband applications of vertical-cavity surface-emitting lasers, wireless sensor body area networks, and emerging optical and wireless technologies for human-to-machine applications. She has coauthored more than 150 journal and conference publications. She is an Optica Fellow and has served on the editorial board of the \emph{Journal of Lightwave Technology} and the \emph{Journal of Optical Communications and Networking}.
\end{IEEEbiographynophoto}

\vfill

\end{document}